\documentclass[useAMS,usenatbib,usegraphicx]{mn2e}
\usepackage[total={17.8cm,24.0cm},centering]{geometry}
\usepackage[british]{babel}
\usepackage{times}
\usepackage{psfig,epsfig}
\newcounter{subfigure}

\title[Two-dimensional stellar and emission-line kinematics of 18 late-type spirals]{Late-type galaxies observed with {\bf SAURON}. Two-dimensional stellar and emission-line kinematics of 18 spirals}

\author[Ganda et al.]{Katia Ganda,$^1$\thanks{E-mail: katia@astro.rug.nl}
Jes\'us Falc\'on-Barroso,$^2$
Reynier F.\ Peletier,$^1$
Michele \ Cappellari,$^2$
\newauthor Eric Emsellem,$^3$
Richard M.\ McDermid,$^2$
P.~Tim de Zeeuw,$^2$
and C. Marcella Carollo$^4$\\
$^1$Kapteyn Astronomical Institute, Postbus 800, 9700 AV Groningen,The Netherlands\\
$^2$Leiden Observatory, Postbus 9513, 2300 RA Leiden, The Netherlands\\
$^3$Centre de Recherche Astrophysique de Lyon - Observatoire, 9~Avenue Charles Andr\'e, 69230 Saint Genis Laval, France\\
$^4$Eidgen\"ossische Technische Hochschule Zurich, H\"onggerberg HPF G4.3, CH-8092 Zurich, Switzerland}

\date{Released 2005 Xxxxx XX}

\pagerange{\pageref{firstpage}--\pageref{lastpage}} \pubyear{2005}

\def\LaTeX{L\kern-.36em\raise.3ex\hbox{a}\kern-.15em
    T\kern-.1667em\lower.7ex\hbox{E}\kern-.125emX}

\begin{document}
\label{firstpage}
\maketitle

\begin{abstract}
 We present the stellar and gas kinematics of a sample of 18 nearby late-type spiral galaxies (Hubble types ranging from Sb to Sd), 
 observed with the integral-field spectrograph {\tt SAURON} at the 4.2-m William Herschel Telescope. {\tt SAURON} covers the spectral range 
 4800-5380 \AA, allowing us to measure the H$\beta$, Fe, Mg{\textit{b}} absorption features and the emission in the 
 H$\beta$ line and the [OIII]$\lambda\lambda$4959,5007\AA\, and [NI]$\lambda\lambda$5198,5200\AA\, doublets over a 33$''$$\times$ 41$''$ field of view.  
The maps cover the
 nuclear region of these late-type galaxies and  in all cases include the entire bulge. In many cases the stellar kinematics suggests the presence of a cold inner region, as visible from a central drop 
 in the stellar velocity dispersion. The ionised gas is almost ubiquitous and behaves in a complicated fashion: the gas 
 velocity fields often display more features than the stellar ones, including wiggles in the zero-velocity lines, irregular distributions, 
 ring-like structures. The line ratio [OIII]/H$\beta$ often takes on low
 values over most of the field, probably indicating a wide-spread star formation.  
 \end{abstract}

\begin{keywords}
 galaxies: late-type - galaxies: bulges - galaxies: discs - galaxies: kinematics - galaxies: formation - galaxies: evolution 
\end{keywords}

\section{Introduction}
 From a theoretical point of view, we have a well-defined paradigm for the
formation of disc galaxies within the Cold Dark Matter (CDM) hierarchical
structure formation scenario (\citealt{fall}, \citealt{silk}): discs quietly settle and cool inside dark matter haloes,
while bulges form through mergers of multiple haloes. However, some of the observed
properties of spiral galaxies suggest a larger complexity in their formation
history. The presence of bulges is not ubiquitous and their nature can be ambiguous. Evidence has
accumulated in the past years showing that many bulges have a disc-like, sometimes exponential radial fall-off of 
the stellar density (\citealt{andredakis}, Andredakis, Peletier \& Balcells 1995, \citealt{jong}, Courteau, de  Jong \&
Broeils 1996, Carollo \& Stiavelli 1998, \citealt{seigaretal}, MacArthur, Courteau \& 
Holtzmann 2003). Numerical
simulations seem to suggest that the dissolution of bars inside the discs may trigger the formation of three-dimensional stellar structures with roughly exponential
profiles (\citealt{pfenniger}, \citealt{combes}, \citealt{raha}, Norman, Sellwood, Hasan 1996); this could mean that some bulges form through the
evolution of dynamical instabilities in the disc. Quite recently, the quality of imaging data made available through HST boosted the study of the inner regions of spiral galaxies, showing that they can host a variety of structures: 
bulges, nuclear star clusters, stellar discs, small bars, double bars, star-forming rings (\citealt{marcella97}, \citealt{marcella98b}, Carollo, Stiavelli \& Mack 1998,  
\citealt{marcella99}, \citealt{perez}, \citealt{marcella02}, \citealt{boker}, \citealt{laine}, Falc\'on-Barroso et al. 2005, \citealt{emma}), 
without there being an agreement about their origin and evolutionary pattern. Ongoing large projects like the panchromatic SINGS survey \citep{sings} which makes use of  
observations at infrared, visible and ultraviolet wavelengths represent a very useful approach to building a comprehensive picture of 
galactic structure, but at the moment rely mostly on imaging. Looking at disc galaxies from a spectroscopic perspective would add kinematic information and insight into stellar 
populations which cannot come from imaging, and could help us tracing their star formation and mass assembly histories.\\ 
\indent Contrary to the massive spheroids, the stellar populations and kinematics of late-type disc-dominated galaxies are poorly known, due to the difficulty of reliably
measuring and interpreting such diagnostics in low surface brightness environments which are so full
of dust, star formation and substructures: not much attention has been paid 
to the spectroscopic counterpart of all the mentioned imaging that has been carried out. There are a few exceptions to this statement: \citet{boker01} 
started a project on STIS long-slit spectroscopy of 77 nearby late-type spiral galaxies
previously imaged with HST/WFPC2; first results are discussed in \citet{boker03}; \citet{walcher} analysed UV slit-spectroscopy of the nuclei of nine late-type spirals; these studies are mainly focussed on the nature of the innermost 
components, in particular on the nuclear star clusters.\\
\indent We are currently engaged in a study aimed at investigating the properties of the nuclear regions of very
late-type galaxies. In such environments, long-slit spectra are too limited to be useful for modelling and interpretation and have
generally been used only to discuss the properties of emission-lines (see for example \citealt{gallag}, who measure the position-velocity 
curve of 21 extreme late-type spiral
galaxies using the H$\alpha$ emission-line). Here we present deep integral-field 
spectroscopy that not only makes it easier to study the kinematics and physical properties of stars and gas, but also allows to study
and model the stellar populations.\\
\indent We were granted 6 nights at the
William Herschel Telescope (WHT) of the Observatorio del Roque de los Muchachos in La Palma,
Spain, to obtain two-dimensional spectroscopy with the integral-field
spectrograph {\tt SAURON}, which was custom-built 
for a representative census of elliptical and lenticular galaxies, and Sa bulges (the 
so-called {\tt SAURON} survey, see Bacon et al.\ 2001, de Zeeuw et al.\ 2002,
hereafter, respectively, Paper I, Paper II). The present work can be regarded as a natural extension of the {\tt SAURON} survey towards
the end of the Hubble sequence. Our purpose was to use {\tt SAURON} in order to map the stellar and gaseous (H$\beta$, [OIII], [NI])
kinematics and the absorption line-strength distributions of the indices H$\beta$, Mg, Fe, in the region 4800-5380\AA. In this paper we present the observations and data reduction and the
resulting kinematical maps for 18 Sb-Sd galaxies. The data
and maps will be made available via the {\tt SAURON} website
(http://www.strw.leidenuniv.nl/sauron\,).\\
\indent The paper is structured as follows. Section \ref{samplesec} describes the sample selection and characteristics; 
Section \ref{observationsec} summarizes the observations and data reduction;
Section \ref{methodsec} describes the methods applied to calculate the stellar and gaseous kinematics from our spectra; 
Section \ref{comparison} carries out a comparison with previous measurements; 
Section \ref{mapssec} presents and discuss the kinematical maps and looks in particular at the behaviour of the
stellar velocity dispersion. Finally, Section \ref{conclusionsec} summarizes the results. 
Detailed modelling and interpretation of the data will come in future papers. 

\section{THE SAMPLE}\label{samplesec}
\begin{center}
\begin{figure*}
{\includegraphics[width=0.99 \linewidth]{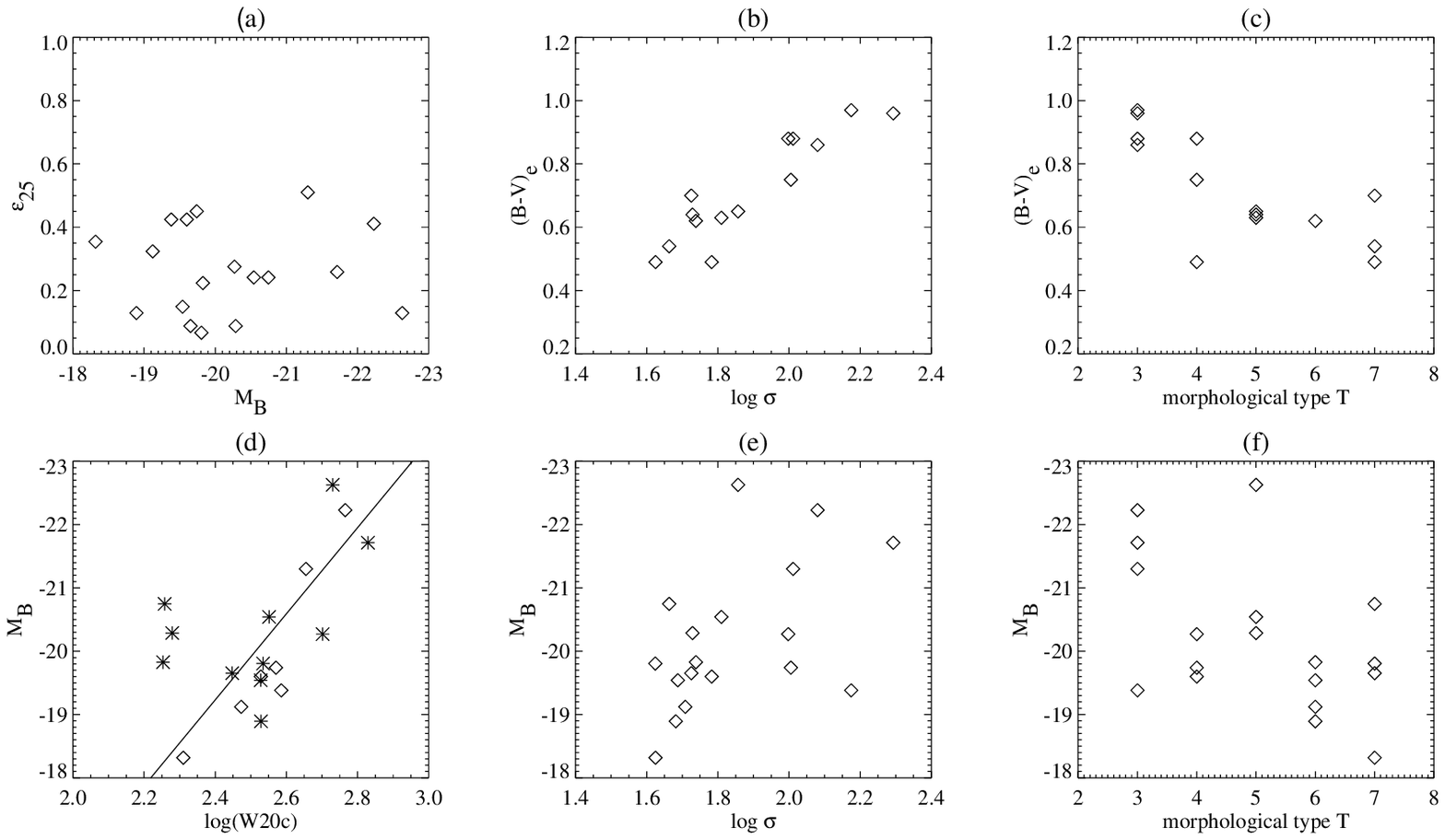}}
\caption{Panel (a): distribution of the 18 spiral galaxies in the plane ellipticity $\epsilon_{25}$ - absolute blue magnitude $M_{B}$; panel (b): 
effective $B - V$ colour versus central velocity dispersion $\sigma$ (in logarithmic units; we note here once for all that through the paper we will 
refer to decimal logarithms simply as logarithms); panel (c): effective 
 $B - V$ colour versus morphological type; panel (d): absolute magnitude $M_{B}$ versus the inclination-corrected rotation velocity (in logarithmic units); 
 the overplotted solid line indicates the standard Tully-Fisher relation (see the text for further details); panel (e): $M_{B}$ plotted versus central velocity dispersion $\sigma$ (in logarithmic units); panel (f): 
$M_{B}$ versus morphological type. The numerical values 
for the plotted quantities are derived from public catalogues (see Table \ref{properties}), except the values for $\sigma$, that come from our own measurements, 
as explained in the text.}
\label{sampleprop}
\end{figure*}
\end{center}

Our sample galaxies were optically selected ($B_{T}$ $<$ 12.5, according to the values given in \citealt{RC3},
hereafter RC3) with
HST imaging available from WFPC2 and/or NICMOS. In practise, the galaxies were chosen from objects lists of recent imaging projects with HST (\citealt{marcella97}, 1998, 2002, 
\citealt{laine}, \citealt{boker}). Their morphological type ranges
between Sb and Sd, following the classification reported in NED\footnote{http://nedwww.ipac.caltech.edu} (from the RC3). Galaxies in close
interaction and Seyferts were discarded. Only galaxies with $0<$RA$<15h$ and $\delta > -20^\circ $
were selected, to fulfill a visibility criterion during the allocated nights. The resulting sample contains 18 nearby galaxies.\\
\begin{table*}
\begin{center} 
 \begin{tabular}{@{}  l l l l l c l l l c l  l l}
  \hline \hline
NGC &Type &T & V$_{21}$ & $M_{B}$& ($B$-$V$)$_{e}$& $d_{25}$&$R_{e}$ & $\epsilon_{25}$&$i$ & PA& $\sigma$ & W${_{20}}_{c}$\\
(1) & (2) & (3) & (4) &   (5) & (6) &   (7) & (8) &(9) & (10) &(11) & (12) &(13)\\
\hline
488& SA(r)b & 3.0 & 2269 & -21.71 & 0.96 & 315 & 52 &0.260 & 42 & 15 & 196& 675\\
628& SA(s)c & 5.0 & 656 & -20.29 & 0.64 & 629 &144 & 0.088& 24&25 & 54& 190\\
772& SA(s)b & 3.0 & 2458  & -22.23 & 0.86 & 435 &77 &0.411 & 54& 130&120&583\\
864& SAB(rs)c & 5.0  & 1560  & -20.54 & 0.63 &281 & 97 &0.241 & 41 &20&65&356\\
1042& SAB(rs)cd & 6.0& 1373&-19.83 & 0.62 & 281&95 & 0.224 & 39& 43$^{g}$&55&179\\
2805& SAB(rs)d & 7.0 & 1734 & -20.75 & 0.54 & 379 &128 & 0.241 &41&125&46&181\\
2964& SAB(r)bc & 4.0 &  1321& -19.74 & 0.75 &  173 & 26 & 0.451 &57&97  & 101&372\\
3346& SB(rs)cd & 6.0 & 1260 &-18.89 &\,\,& 173 && 0.129 &29&111$^{g}$&48&338\\
3423& SA(s)cd & 6.0 &  1011 & -19.54 & & 228 &36 & 0.149 &32&10  &49&337\\
3949& SA(s)bc & 4.0 &  798& -19.60 &\,\, 0.49$^{L}$ & 173& & 0.425 &55&120&61&338\\
4030& SA(s)bc & 4.0  & 1460 & -20.27 & \,\,\,0.88$^{L}$ & 250 & &0.276 & 44&27&100&503\\
4102& SAB(s)b? & 3.0 & 837& -19.38 &\,\, 0.97$^{L}$ & 181 && 0.425 & 55&38&150&385\\
4254& SA(s)c & 5.0  & 2407& -22.63 & 0.65 & 322 &56 &0.129 & 29&62$^{g}$ &72&537\\
4487& SAB(rs)cd & 6.0  & 1037&  -19.12 & & 250 && 0.324 & 47&75&51&297\\
4775& SA(s)d & 7.0 & 1567& -19.81 & & 128 && 0.067 & 21& 52$^{g}$&42&342\\
5585& SAB(s)d & 7.0 & 305&   -18.32 & 0.49 &345 &102 &0.354 & 50 &30 &42&204\\
5668& SA(s)d & 7.0&1583& -19.65 &0.70& 199 & 38 & 0.088 &24&164$^{g}$ &53&280\\
5678& SAB(rs)b & 3.0 & 1922& -21.30 &\,\, 0.88$^{L}$ &199 &  &0.510 & 61&5&103&452\\
\hline
 \end{tabular}\\
\caption{Properties of our 18 galaxies. (1) Galaxy identifier. (2) Hubble type (RC3 through NED). (3) Numerical morphological type (RC3).
 (4) Heliocentric neutral hydrogen velocity in km s$^{-1}$
 (RC3, via VizieR).
  (5) Absolute blue magnitude $M_{B}$ in mag (quoted from HyperLeda), computed from the corrected apparent magnitude and the distance
  modulus (also listed in HyperLeda); a Virgocentric flow model with $v_{virgo}$ = 208 km s$^{-1}$, an Hubble constant $H_{0}$ = 
  70  km s$^{-1}$ Mpc$^{-1}$ and the correction to the Local Group centroid of \citet{locgr} are adopted. (6) Effective $(B - V)_{e}$ colour in mag; values  
  marked with ``$L$'' have been taken from HyperLeda, the others from RC3. 
  (7) Projected diameter at the isophotal level of 25 mag arcsec$^{-2}$ in the $B$ band, in
  arcsec (RC3). (8) Effective radius $R_{e}$ in the $B$ band, in arcsec (RC3). 
  (9) Ellipticity $\epsilon_{25}$ of the contour of 25 mag arcsec$^{-2}$ $B$ surface brightness (RC3). (10)
  Disc inclination in degrees, calculated from the axis ratio listed in RC3. (11) Position angle of the major axis, in degrees (RC3);
  values marked with ``$g$'' are taken from \citet{grosbol}.
   (12) Central velocity dispersion $\sigma$ in km s$^{-1}$, from our own measurement (averaging the spectra in a central 2\farcs4 $\times$ 2\farcs4 aperture and measuring the kinematics on 
   the resulting spectrum, as explained in the text). (13) Width of the 21-cm neutral hydrogen line at 20 percent of
   the peak in km s$^{-1}$ (RC3), corrected for the disc inclination, as listed in column (10). }
\label{properties}
\end{center}
\end{table*} 
\indent In Table \ref{properties}
we list properties already measured and available through public catalogues, while in Figure \ref{sampleprop} we represent graphically the range 
spanned by our sample galaxies in a number of global and nuclear properties. This can be useful for a  visual comparison with the galaxies of the {\tt
SAURON} survey (see Figure 1 and Figure 3 of \citet{paper2}). Panel (a) shows the
distribution of the selected galaxies in the plane $M_{B}$ - $\epsilon_{25}$. $M_{B}$ is the absolute magnitude in the $B$-band and $\epsilon_{25}$ is 
the ellipticity, derived from 
the axial ratio at the 25 mag arcsec$^{-2}$ isophotal level in $B$. This panel shows that there is a lack of high-ellipticity objects,  
indicating that our disc galaxies are generally far from being edge-on systems. Panel (b) 
plots the effective $B - V$ colour versus the central velocity dispersion $\sigma$; a relatively tight trend is recognizable, since galaxies with 
higher velocity dispersions tend to be redder. Panel (c) plots the 
effective $B - V$ colour versus the morphological type; colours become bluer with later types.                                       
Panel (d) shows the distribution of our galaxies in the Tully-Fisher plane: $M_{B}$ versus the inclination-corrected rotation velocity, obtained as 
$W_{20_{c}}=W_{20}/\sin(i)$, where $W_{20}$ is the 21-cm line width at 20
percent of the peak and $i$ the inclination between line of sight and polar axis. The overplotted solid line is the Tully-Fisher relation, as
determined by \citet{marc2001} for the $B$ band from a sample of 45 galaxies with measured HI global profile: 
$M_{B} = -2.91 - 6.8  \log(W_{20_{c}})$. In the Figure, all the galaxies with inclination below 
$45^\circ$ are marked with an asterisk, since the reported relation was established on the basis of higher inclination objects: one can see that all the galaxies 
that deviate most from the red line have low-inclination, although not all of the low-inclination galaxies are deviant. Panel (e) presents the relation between total luminosity (absolute blue magnitude $M_{B}$) and 
central velocity dispersion; the galaxies cover a range of $\approx 100$ in
luminosity and a factor $\approx$ 5 in velocity dispersion and become more luminous with increasing velocity dispersion. To conclude, panel (f) plots $M_{B}$ against the morphological type; 
the luminosity tends to decrease as the galaxies become later in type, as shown also by \citet{jong96}. 
The quantities plotted in this Figure are all derived from catalogues, mainly from RC3 and HyperLeda\footnote{http://leda.univ-lyon1.fr}, 
(for a description and a list of references, see caption to Table \ref{properties}), 
except for the central velocity dispersion.
Since previously existing central velocity dispersion values are available only for six of the galaxies in our sample, we decided to measure the central velocity dispersion from our own spectra. 
The measured spectra in a central aperture (2\farcs4 $\times$ 2\farcs4) were combined to give an averaged spectrum from which we computed the stellar kinematics, 
using methods which we will present and discuss later in this paper. The resulting $\sigma$ values are taken as central velocity dispersions,
listed in Table \ref{properties} and used in Figure \ref{sampleprop}. In Section \ref{comparison} 
we will provide a qualitative comparison between our central velocity dispersions and the literature values, 
for the six galaxies for which we could find references.

\section{Observations and data reduction}\label{observationsec}
\begin{figure*}
{\includegraphics[width=0.85\linewidth]{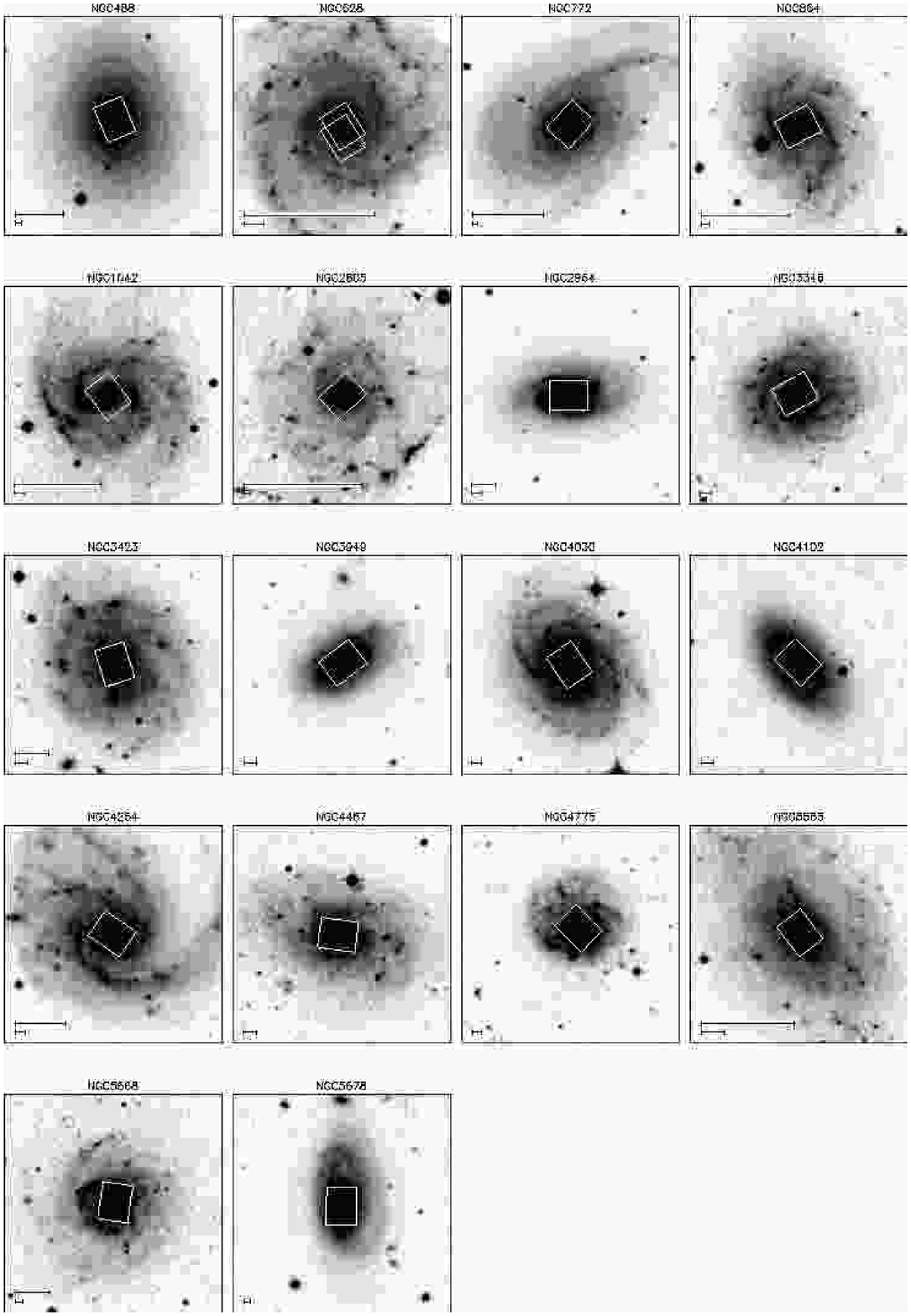}}
\caption{$R$-band Digital Sky Survey images of all 18 late-type spirals in the sample. The size of each image is 4$'$ $\times$
4$'$ and the orientation is such that North is up and
East is left. Overplotted on each image are the positions of the {\tt SAURON} pointings. The small bar at the bottom-left corner of each image 
corresponds to the linear length of 1 kpc; when the effective radius is available from RC3 (Table
\ref{properties}), another bar indicating the size of 1 $R_{e}$ is provided, above the 1 kpc bar.}
\label{sample}
\end{figure*}

Observations of the 18 late-type galaxies were carried out on January 20-26, 2004 using the
integral-field spectrograph {\tt SAURON} attached to the 4.2-m William Herschel
Telescope (WHT). For each of the 18 galaxies, Table \ref{pointings} lists the number of 1800s exposures 
that we took.\\
\indent We used the low spatial resolution mode of {\tt SAURON}, giving a field-of view
(FoV) of 33$''$ $\times$ 41$''$. The spatial sampling of individual exposures is
determined by an array of 0\farcs94 $\times$ 0\farcs94 square lenses. This produces 1431
spectra per pointing over the {\tt SAURON} FoV; another 146 lenses sample a region 1\farcm9 away
from the main field in order to measure simultaneously the sky background. 
{\tt SAURON} delivers a spectral resolution of 4.2 \AA\, FWHM
and covers the narrow spectral range 4800-5380 \AA\, (1.1 \AA\, per pixel). This
wavelength range includes a number of important stellar absorption lines (e.g.
H$\beta$, Fe, Mg{\textit{b}}) and potential emission-lines as well 
(H$\beta$, [OIII], [NI]). For a more exhaustive description of the instrument, see \citet{paper1} and in particular 
Table 1 there.\\
\indent For each galaxy, two to six largely overlapping exposures of 1800s were typically
obtained (Table \ref{pointings}). An offset of a few arcseconds, which corresponds to a few spatial
elements, was introduced between consecutive exposures to avoid systematic errors due for example to bad CCD regions. Figure \ref{sample} outlines the (approximate) position of the {\tt SAURON} pointings 
overlaid on $R$-band Digital Sky Survey images of our galaxies, showing that our observations cover the nuclear regions.
\begin{table}
 \begin{center}
 \begin{tabular}{c c c c c   }
  \hline\hline
  NGC & \# &  &NGC & \# \\
  \hline
  488&3  & &3949&3 \\
  628&1+4$^{a}$  & &4030&5 \\
  772&5  & &4102&5 \\
  864&4  & &4254&2 \\
  1042&4  & &4487&4 \\
  2805&6  & &4775&4 \\
  2964&2  & &5585&5 \\
  3346&6  & &5668&5 \\
  3423&6  & &5678&4 \\ 
  \hline
  \end{tabular}
 \caption{Number of exposures of 1800 s each per galaxy. Note $a$: for the first exposure of NGC628 we
 pointed the telescope on a star offset by $\approx$ 13$''$ with respect to the galaxy centre; so we have 2 slightly
 different pointings on this galaxy.}
\label{pointings} 
\end{center}
\end{table}

\subsection{Data reduction}
We reduced the {\tt SAURON} observations using the dedicated software {\tt XSAURON} developed at CRAL \citep{paper1}. During the observing run, arc exposures were taken before
and after each galaxy exposure for wavelength calibration purposes. Tungsten lamp exposures were also taken every night in order to build the extraction mask. 
At the telescope we
had a misalignment of $\approx 1^\circ$ between the columns of the CCD and the dispersion direction. To correct for this misalignment and avoid interference patterns due
to uneven sampling of the data, we decided to rectify the spectra by rotating all of the frames by the same amount ($\approx 1^\circ$) and in the opposite sense to the mentioned misalignment, at a very early stage of the
 reduction, by means of the IRAF\footnote{IRAF is distributed by the National Optical Astronomy Observatories, which are operated by the Association of Universities for 
 Research in Astronomy, Inc., under cooperative agreement with the National
 Science Foundation.} tasks {\tt geomap} and {\tt geotran}, available from the {\tt images.immatch} package. The reduction 
steps include thus bias and dark subtraction, rotation of all the frames,  
extraction of the spectra using the fitted mask model, wavelength calibration, low-frequency flat-fielding, cosmic-ray removal, homogeneization of the spectral resolution
over the field of view, sky subtraction and flux 
calibration of the spectra, although the data were not necessarily collected under 
photometric conditions. The individually extracted and calibrated 
datacubes were finally merged by truncating the 
wavelength domain to
a common range, recentering the exposures using 
reconstructed images and combining the spectra, 
while correcting also for the effect of atmospheric 
refraction. In this process, the datacubes were
spatially resampled to a common grid, so that  the 
final merged datacube is sampled on to a rectangular 
grid with 0\farcs8 $\times$ 0\farcs8 pixels. The
improvement in spatial sampling with respect to the individual datacubes is due to the dithering of exposures.

\section{ANALYSIS AND METHODS}\label{methodsec} 
In order to ensure the measurement of reliable stellar kinematics, we spatially binned our merged datacubes using the Voronoi 2D binning algorithm of \citet{voronoi}, 
creating compact bins with a minimum signal-to-noise ratio $(S/N)_\star \approx$ 60 per resolution element. However, most of the spectra in the central regions have a
$(S/N)_\star$ greater than 60, so a large fraction of the original spatial elements remains unbinned. 

\subsection{Stellar kinematics}\label{methodstar}
 We measured the stellar kinematics on each spectrum in our binned datacubes using the penalized pixel fitting (pPXF) method by \citet{ppxf}. A linear 
 combination of template stellar spectra, convolved with a line-of-sight velocity distribution (LOSVD) described as a
 Gauss-Hermite expansion (\citealt{vdm}, \citealt{gerhard}), is fitted to each galaxy spectrum by $\chi^{2}$
 minimization in pixel space, using a penalty term to suppress noise. While
 fitting, the spectral regions that are potentially affected by nebular emission (corresponding to the H$\beta$, [OIII], [NI] lines) are masked out. A low-order
 polynomial (generally of order six) is also included in the fit to account for small differences in the flux calibration between the galaxy and the template spectra. 
 This allows us to derive the mean velocity ($V$), velocity dispersion ($\sigma$) and the higher order Gauss-Hermite moments ($h_{3}$ and $h_{4}$).
 As stellar templates we used a library of single-age, single-metallicity population
 models (SSP) from \citet{vazdekis}, from which we selected 39 models characterized by  
1.00 $\leq$ Age $\leq$ 17.78 Gyr, -1.68 $\leq$ [Fe/H] $\leq$ +0.20. This is similar
to what has been done by Falc\'on-Barroso et al. (2005, hereafter Paper VII) for the analysis of the 24 Sa galaxies part of the {\tt SAURON} survey.\\
\indent As we will show in Section \ref{mapssec}, our galaxies display stellar velocity dispersions often lower than those measured in the early-type galaxies of the 
{\tt SAURON} survey (Emsellem et al. 2004, hereafter Paper III). {\tt SAURON} has an instrumental dispersion
of 108 km s$^{-1}$, while our measured velocity dispersions are in many cases below that level (see the 
central values for $\sigma$ listed in Table \ref{properties}). Thus, one might be concerned that velocity dispersions significantly below the
instrumental dispersion cannot be reliably measured . This issue has already been addressed in \citet{paper3}: it reports tests of the uncertainties on the measured $\sigma$ via
Monte-Carlo simulations which prove that for a spectrum with $(S/N)_\star \approx$ 60 and $\sigma \approx$ 50 km s$^{-1}$, the pPXF method will output velocity 
dispersions differing from the intrinsic one by no more than 10 km s$^{-1}$, a value within the measured error.\\
\indent Another technical issue related to the low velocity dispersion values is that when the observed velocity dispersion is less than about 2 pixels 
($\sigma \leq$ 120 km s$^{-1}$) it becomes more difficult to measure the Gauss-Hermite moments $h_{3}$ and $h_{4}$ at our minimum $(S/N)_\star \approx$ 60; 
in those cases the penalization in pPXF then biases the solutions towards a Gaussian \citep{ppxf}. In practise one expects the measured higher moments to be significant only in the galactic regions 
where the velocity dispersion or $(S/N)_\star$ are high.

\subsection{Gas kinematics}\label{methodgas}
\begin{center}
\begin{figure*}
{\includegraphics[width=0.99\linewidth]{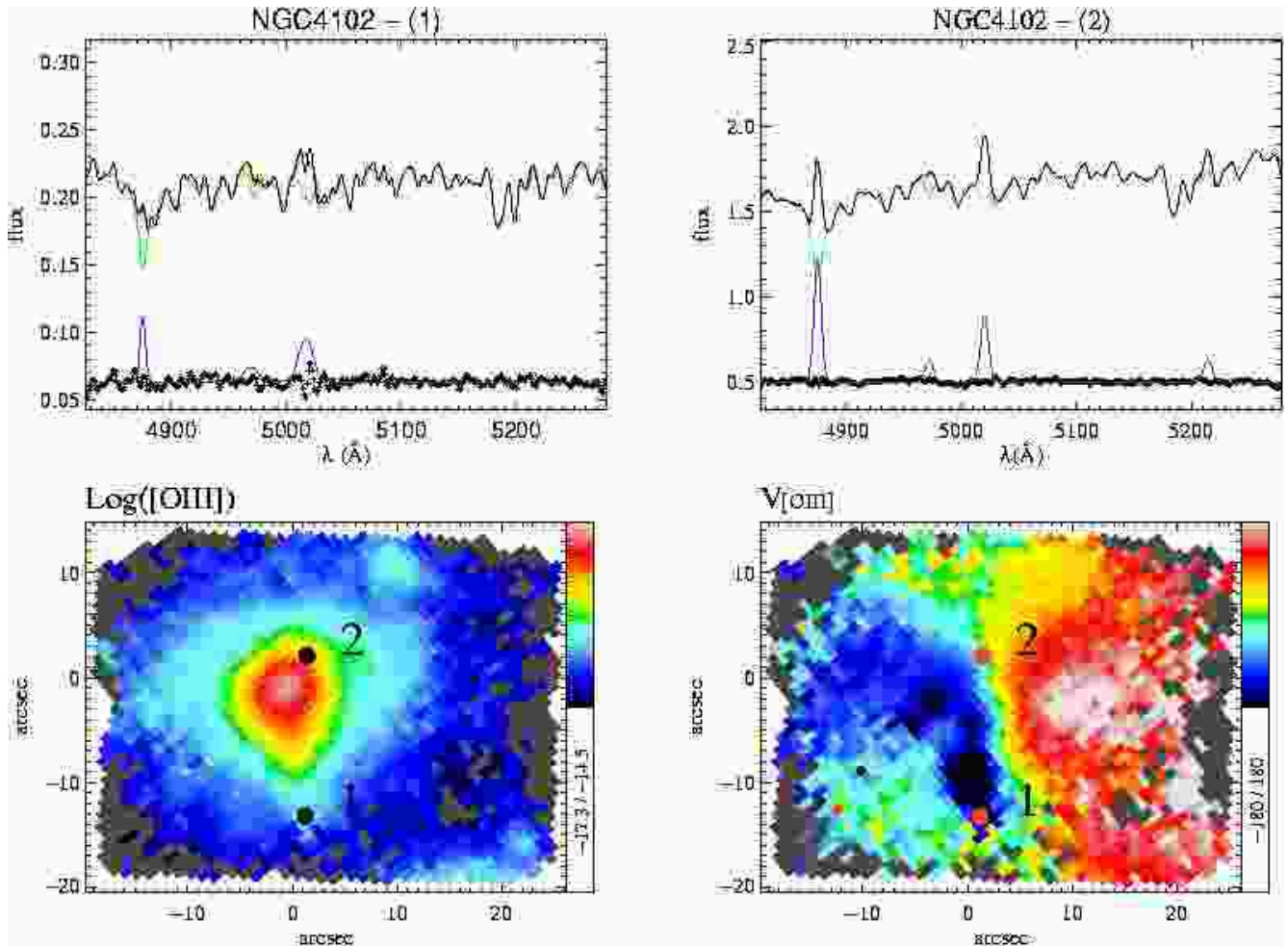}}
\caption{In the top left panel we show for NGC4102 a spectrum where the [OIII] lines are only badly fitted by a single 
Gaussian; in the plot, the black solid line represents the galaxy spectrum; the red line is the full spectrum fit; the green solid
line is the best-fitting combination of stellar templates, convolved with the kinematics; at the bottom of the panel, the purple line is the pure-emission spectrum
and the black dotted line represents the residuals from the fit; a constant has been added to both. Along the vertical axis, the flux 
is in units of 10$^{-16}$ erg cm$^{-2}$ s$^{-1}$. The top right panel shows a spectrum for the same galaxy where there are no deviations from Gaussianity in the line profiles. This is an example of the typical behaviour of our spectra and 
also illustrates the quality of fit that we obtain. In the second row the [OIII] flux (in erg cm$^{-2}$ s$^{-1}$ and logarithmic scale) and velocity (in km s$^{-1}$) maps are shown (bottom left and right panels respectively); 
the location of the two representative spectra is marked on the maps with a dot. For a description of the maps, we refer the reader to Section 6.1.}
\label{bin_bad}
\end{figure*}
\end{center}
Nebular emission is almost ubiquitous in our sample of 18 spiral galaxies. Thus we have decided to apply a ``double-binning scheme'', in order to compute the gas kinematics on bins
smaller than the ones with $(S/N)_\star = 60 $ on which we calculated the stellar kinematics as described above. The large amount of gas in fact allows us to reliably 
measure the kinematics of emission-lines at a lower $S/N$ level, and having smaller bins prevents loss of spatial resolution, especially in the outer regions. To do this, we
proceed as follows.\\
\indent We call ``star-bins'' the previously introduced Voronoi-bins with minimum $(S/N)_\star = 60$. We first reconstruct the gas map by running pPXF on each single unbinned spectrum (single-lens spectrum) 
keeping fixed the stellar kinematics at the values determined for the star-bin to which the spectrum
belongs, thus fitting only for the stellar template and polynomial coefficients; this is done again by masking out the spectral regions potentially affected by emission. 
We recover the gas spectrum by subtracting the best-fit from the original spectrum. For each gas spectrum we compute the $(S/N)_{gas}$ of the [OIII] line taking the peak of 
the considered gas spectrum over the [OIII] region and dividing by the noise, calculated on the gas spectrum over the emission-free part of the {\tt SAURON} range.
For the $(S/N)_{gas}$ determination we could have used one of the other emission-lines in the {\tt SAURON} spectral range, e.g. H$\beta$ or [NI], but they both present disadvantages compared to [OIII]. 
In fact, the
lines of the [NI] doublet are normally quite weak, and it is difficult to measure their kinematics and amplitude; the H$\beta$ emission can instead 
be contaminated by stellar absorption at the same wavelength. At this stage, we bin the merged datacube to a minimum $(S/N)_{gas}$ of 5, according to the $(S/N)_{gas}$ 
values found for each single gas spectrum. We end up with a ``gas-binned'' datacube.
We call these bins ``gas-bins''.\\
\indent In a second step, we determine the gas kinematics for each gas-bin. We proceed by running pPXF on the average of all the star-bins 
spectra that have some spatial intersection
with the considered gas-bin. This is the input stellar kinematics for our determination of the gas kinematics.\\
\indent The actual computation of the gas kinematics parameters requires a careful separation of the line emission from the stellar absorption. 
We follow the procedure described in Sarzi et al. 2005 (hereafter Paper V) and in Paper VII, that has been fully tested on the {\tt SAURON} survey galaxies. For each gas-bin spectrum, 
the method relies on an iterative search for the emission-lines velocities and 
velocity dispersions, while linearly
solving at each step for the line amplitudes and the optimal combination of the stellar templates. The fit is performed over each ``gas-binned'' spectrum, and no masking is applied, 
in contrast to the stellar kinematics determination. 
A multiplicative Legendre polynomial of order six is included in the fit to correct for small differences in the flux calibration between the galaxy spectrum
and the library of models.\\
\indent Most of our galaxies have conspicuous gas emission, so we proceed by fitting the H$\beta$ and [OIII] lines independently, in order to detect 
differences in the kinematics of the two lines, if present. The [NI] lines are instead forced to share the same kinematics with H$\beta$
(see \citealt{paper5} and \citealt{jesus} for details on the method).\\  
\indent We have visually inspected the emission-line profiles in our datacubes in order to assess the applicability of 
our fitting method, which fits a single Gaussian profile to each
line in the {\tt SAURON} wavelength range. In the large majority of cases, we did not find deviations from pure 
Gaussians. Only NGC2964 and NGC4102 present complex line profiles, but limited to specific regions possibly related 
to activity (see the description of individual objects in Section \ref{notes}). This lack of complex line profiles 
may well be caused by the limited instrumental spectral resolution. For completeness, in Figure 
\ref{bin_bad} we show for NGC4102 a spectrum where the [OIII] 5007 \AA\, line does not resemble a 
single Gaussian and a spectrum where the line profiles instead do not 
differ from the Gaussian fit; there we present also the [OIII] flux and velocity maps, in order to spatially locate the mentioned spectra.

\section{COMPARISON WITH PUBLISHED MEASUREMENTS}\label{comparison}
\begin{figure*}
\includegraphics[width=0.99\linewidth]{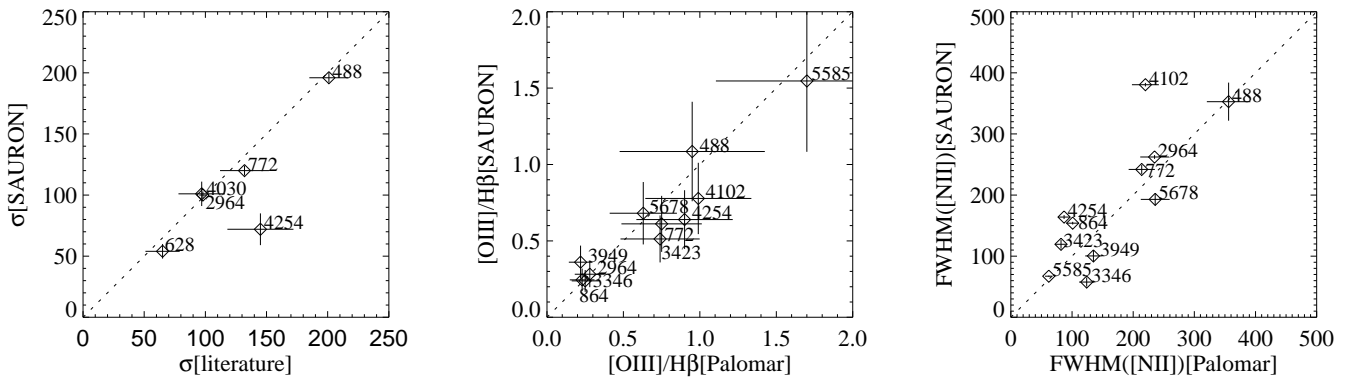}
\caption{Literature comparison for stars (left panel) and gas measurements (central and right panels). Left panel: {\tt SAURON} against literature central stellar velocity dispersion (in km s$^{-1}$); central panel: {\tt SAURON} against
Palomar [OIII]/H$\beta$ line ratio; right panel: {\tt SAURON} against Palomar forbidden line width (in km s$^{-1}$). In each of the plots, the
dotted line marks the 1:1 relation and the NGC galaxy identifiers are indicated close to the corresponding symbol; see the text for further details.}
\label{comparison_fig}
\end{figure*}

 The methods described in the previous sections were used to measure the stellar and gaseous kinematics and the amount 
of gas emission for the 72 galaxies of the {\tt SAURON} survey (\citealt{paper3}, \citealt{paper5}, \citealt{jesus}). The quoted papers 
showed that the methods give results in agreement with previous measurements.\\
\indent We also carried out a direct comparison with previous work. As already mentioned in
Section \ref{samplesec}, measurements of the stellar velocity dispersion are available for one third of our sample. 
In addition, those few references are very heterogeneous and in some cases do not give all the information required 
to perform a careful and systematic comparison. In any case, we can qualitatively investigate the agreement of the 
central aperture velocity dispersion listed in Table \ref{sampleprop} with the literature. The left panel of Figure \ref{comparison_fig} 
plots our measurements against the average of the existing values\footnote{From \citet{shapiro} for NGC4030 and from HyperLeda
for NGC488, NGC628, NGC772, NGC2964 and NGC4254.} (if more than one). Overplotted with a dotted line is the 1:1 relation. Uncertainties in the 
literature values are taken from the references; as for our own measurements, we give an estimate of the errors 
by running pPXF on the single-lens spectra within our aperture and looking at the scatter in the resulting velocity dispersions. The agreement 
is generally satisfactory; the only deviant galaxy is NGC4254, for which the only references we could find date back from the 1980's 
(\citealt{whitmore}, \citealt{tonry}).\\
\indent In the case of the gas measurements, as a source for our literature comparison we choose the Palomar spectroscopic survey of Ho, Filippenko \& Sargent (1995, 1997), 
which includes 11 out of 
our 18 galaxies. We compare the [OIII]/H$\beta$ line ratio and the width of the forbidden emission, represented by the [OIII] and [NII] 6583\AA\, lines  
of the {\tt SAURON} and Palomar samples respectively. To perform a proper comparison, we measured the [OIII]/H$\beta$ line ratio and [OIII] FWHM 
on central spectra obtained extracting from our datacubes a central aperture that matches the size (2$''$ $\times$ 4$''$) and orientation 
of the Palomar long-slit observations. Uncertainties in the [NII] FWHM in the Palomar survey are typically $\approx$ 10\%, except for 
NGC3346 and NGC3949, where \citet{ho97} report only the 3$\sigma$ upper limit; for the [OIII]/H$\beta$ line ratio, 
the quoted uncertainties are around 30-40\%, except the case of NGC488, where the uncertainty is $\approx$ 50\%. For our {\tt SAURON} data, errors on the 
[OIII] line width come from the fitting procedure and vary percentually from galaxy to galaxy, while an upper limit on the uncertainties on the
[OIII]/H$\beta$ line ratio comes from the typical errors on the line fluxes estimate given in Paper V.\\
\indent The central panel of Figure \ref{comparison_fig} plots the {\tt SAURON} against Palomar [OIII]/H$\beta$ line ratio. The agreement appears to be reasonable, 
with all the galaxies lying on the 1:1 relation (indicated by the dotted line in Figure \ref{comparison_fig}) within the error bar, with the exception of NGC3949. The 
right panel compares the FWHMs of the forbidden lines, and the agreement appears to be less satisfactory. In particular, NGC4102 lies well above the 1:1 relation; this is 
due to the broadening of the [OIII] lines in this galaxy, where the single Gaussian used in our method fails to reproduce the line profile, as demonstrated
at the end of the previous Section. We stress the fact that the Palomar slit was oriented in the direction of the galactic region where the lines are double-peaked (see
Section \ref{methodgas} and \ref{ngc4102}): our extracted aperture tests then that region and thus it is not a surprise that the comparison is not very accurate. In fact, a tight correlation between the widths of the 
[OIII] and [NII] lines is expected if the lines are produced in low-density regions \citep{ho97}, and it is not clear that this condition is satisfied in all objects.  

\section{OBSERVED STELLAR AND GAS KINEMATICS}
\label{mapssec}
This section presents our results for the 18 spiral galaxies and points out and briefly describes some interesting features detectable in the maps. We first show the 
maps (Section \ref{allthemaps}), then discuss some properties of the stellar and gaseous 
kinematics of the sample as a whole (Sections \ref{stkinobs} and \ref{gaskinobs}), and in the end give a
detailed description of the individual galaxies (Section \ref{notes}).

\subsection{Stellar and gas flux and kinematics: the maps}\label{allthemaps}
Figures \ref{maps1}--\ref{maps18} below present the flux and kinematical maps of stars and gas for our sample of 18 spiral
galaxies, obtained as explained in Sections \ref{methodstar} and \ref{methodgas}. For each 
object, in the first row we give the unsharp-masked optical image of the galaxy from HST\footnote{We used WFPC2 images taken with the 
F606W filter for NGC488, NGC628, NGC864, NGC1042, NGC2964, NGC3423, NGC3949, NGC4030, NGC4102, NGC4254, NGC4487, NGC5585, NGC5678 
and with the F814W filter for NGC2805, NGC3346, NGC4775, NGC5668.} together with the UGC number when available, the FK5 2000 coordinates,
the absolute blue magnitude, the ellipticity and the morphological type; NGC772 represents an
exception in that respect, since an optical space-based image is not available: we therefore show the unsharp-masked {\tt
SAURON} image instead; the second row
contains the total intensity reconstructed by collapsing the {\tt SAURON} spectra in the wavelength direction (in mag arcsec$^{-2}$, with
an arbitrary zero point), and the stellar velocity and velocity 
dispersion (in km s$^{-1}$); the third and fourth rows contain respectively the H$\beta$  and [OIII] flux (in erg cm$^{-2}$ s$^{-1}$ and logarithmic scale) and kinematics (velocity and velocity dispersion, in km s$^{-1}$); finally, the fifth row 
presents the line ratio [OIII]/H$\beta$, in logarithmic scale, and the stellar $h_{3}$ and $h_{4}$ kinematical moments. Overplotted on each map are the isophotal 
contours. All the plotted velocities are 
systemic velocity subtracted; the same systemic velocity is assumed for stars and gas. No inclination correction to the kinematics has been applied, thus the plotted velocities
are in all cases projected velocities. The stellar kinematics is shown on the $(S/N)_\star = 60$ star-bins, while the gas parameters are plotted on the $(S/N)_{gas} = 5$ gas-bins. 
In each Figure, the arrow and its associated dash at the top of the page, close to the galaxy name, indicate the orientation of the maps, pointing to 
the North and East directions respectively.\\
\indent To display the gas maps, we have chosen to consider as real a detection of emission when the amplitude over noise (hereafter A/N) of the line, defined as the fitted emission amplitude 
divided by the noise in the residual spectrum (galaxy spectrum $-$ best-fit over the whole {\tt SAURON} range), is larger than
4 (see also \cite{paper5}, \cite{jesus}). The bins below this threshold are displayed using a dark grey colour. Despite this cut in A/N, the gas detection covers 
in most cases a very large fraction of the {\tt SAURON} field.\\
\indent In NGC4030 and NGC4102 we detected also significant emission from the [NI] 
doublet; however, in the large majority of our galaxies [NI] is very weak and hard to measure: since we would not learn more about our galaxies considering also the [NI] maps,
we decided not to include them in Figures \ref{maps1}-\ref{maps18}.

%
% Begin maps figure. Use 'subfigure' counter (defined in preamble)
% to give Caption 

\renewcommand{\thefigure}{\arabic{figure}\alph{subfigure}}
\setcounter{figure}{4}
\setcounter{subfigure}{1}
%ngc488
\clearpage
\begin{figure*}
\begin{center}
{\includegraphics[width=0.99\linewidth]{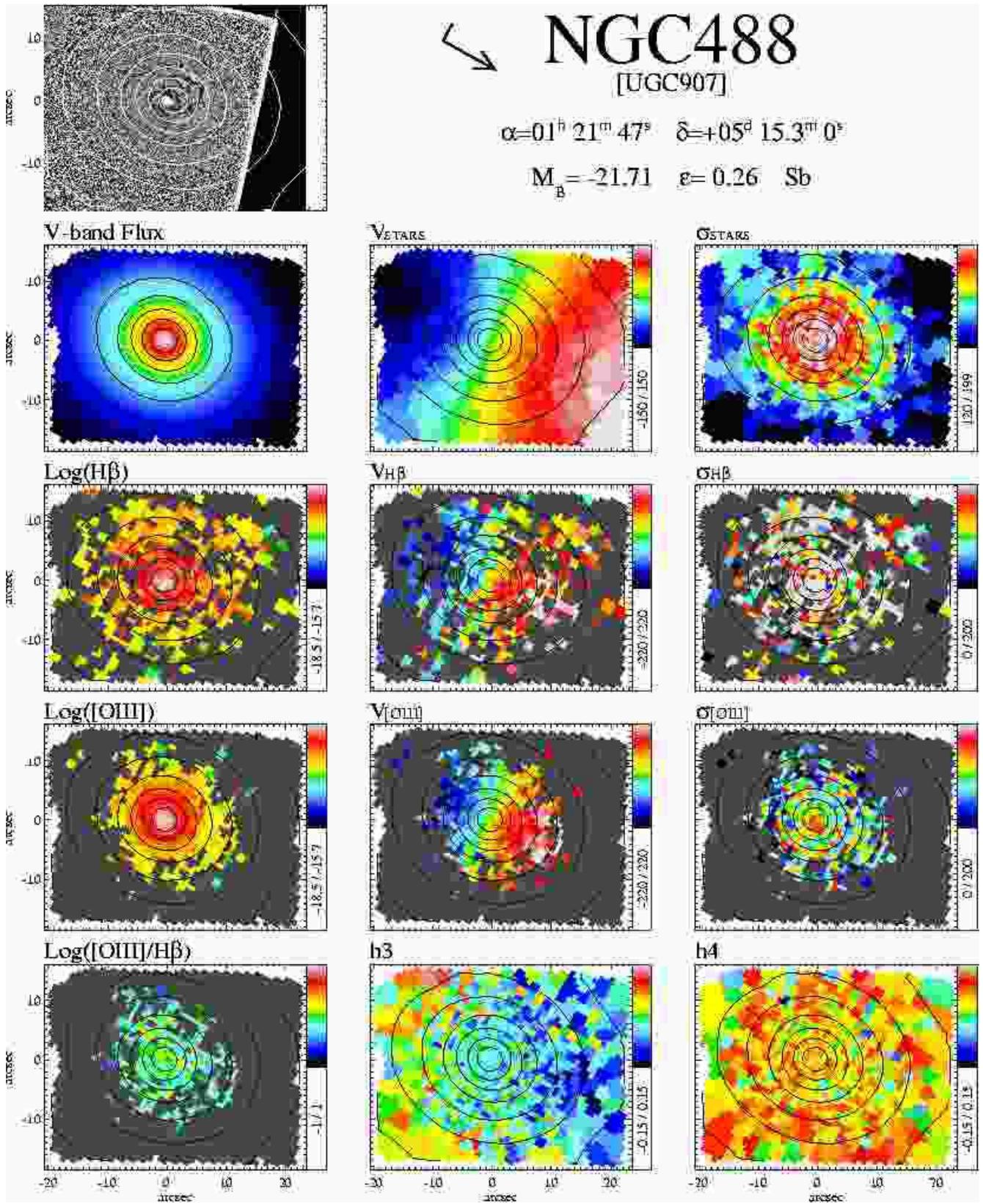}}
\end{center}
\caption{Maps of stellar and gaseous kinematics for NGC488. First row:
unsharp-masked image and some astrometric
information; second row: stellar flux (in mag arcsec$^{-2}$, with an arbitrary zero point), velocity and velocity dispersion (in km s$^{-1}$); third and fourth row: respectively H$\beta$ and [OIII] flux
(in erg s$^{-1}$ cm$^{-2}$ and logarithmic scale), velocity and velocity dispersion (in km s$^{-1}$); fifth
row: [OIII]/H$\beta$ line ratio (in logarithmic scale), stellar $h_{3}$ and $h_{4}$ moments. The ranges are indicated in the box on the right of each
map. In the gas maps, the dark grey colour is used 
for the bins with A/N below the selected threshold.}
\label{maps1}
\end{figure*}

% Have to keep figure counter at previous value, and progress the
% character counter:
\addtocounter{figure}{-1}
\addtocounter{subfigure}{1}

%ngc628
\clearpage
\begin{figure*}
\begin{center}
{\includegraphics[width=0.99\linewidth]{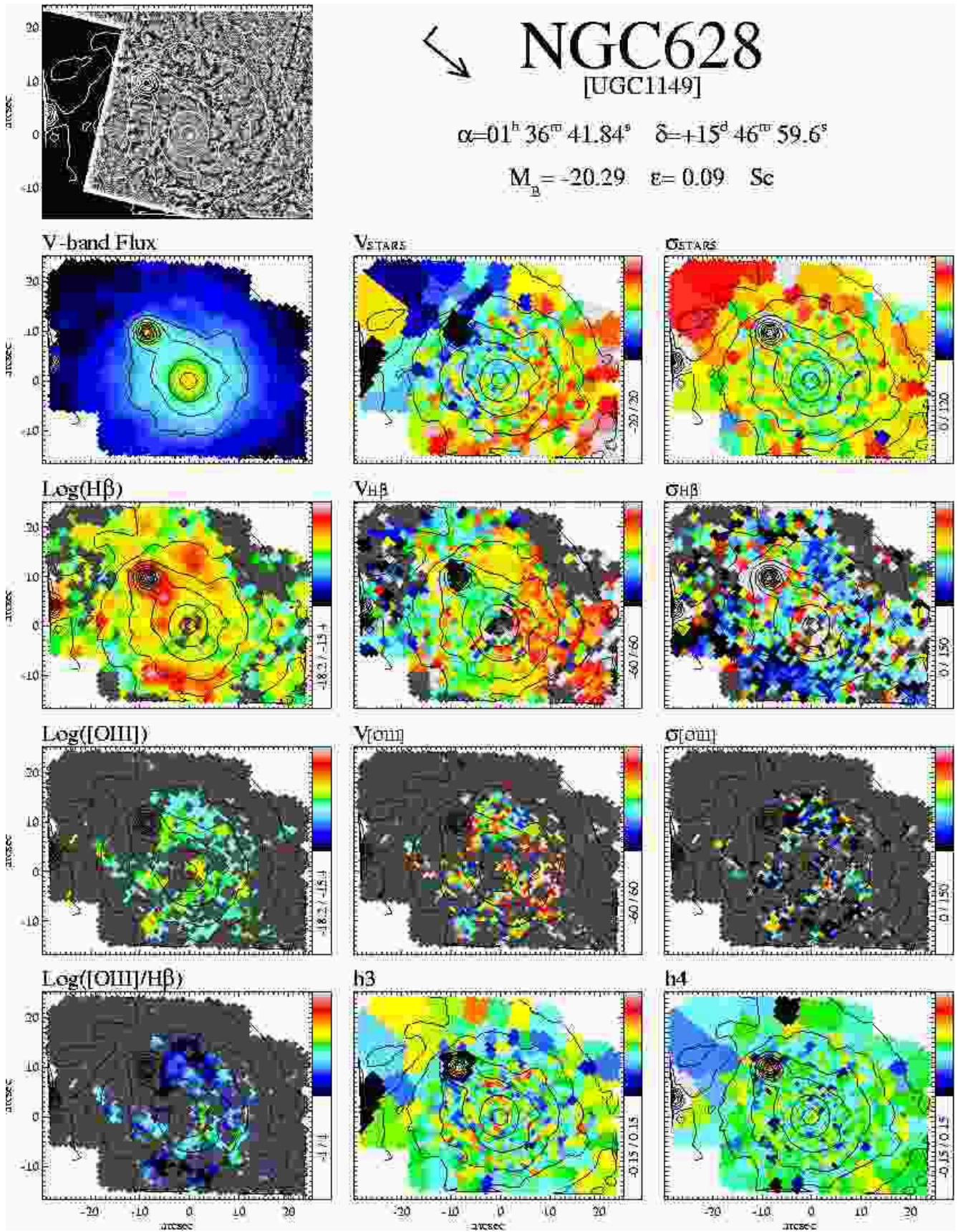}}
\end{center}
\caption{As in Figure \ref{maps1} for NGC628.}\label{maps2}
\end{figure*}
\addtocounter{figure}{-1}
\addtocounter{subfigure}{1}

%ngc772
\clearpage
\begin{figure*}
\begin{center}
{\includegraphics[width=0.99\linewidth]{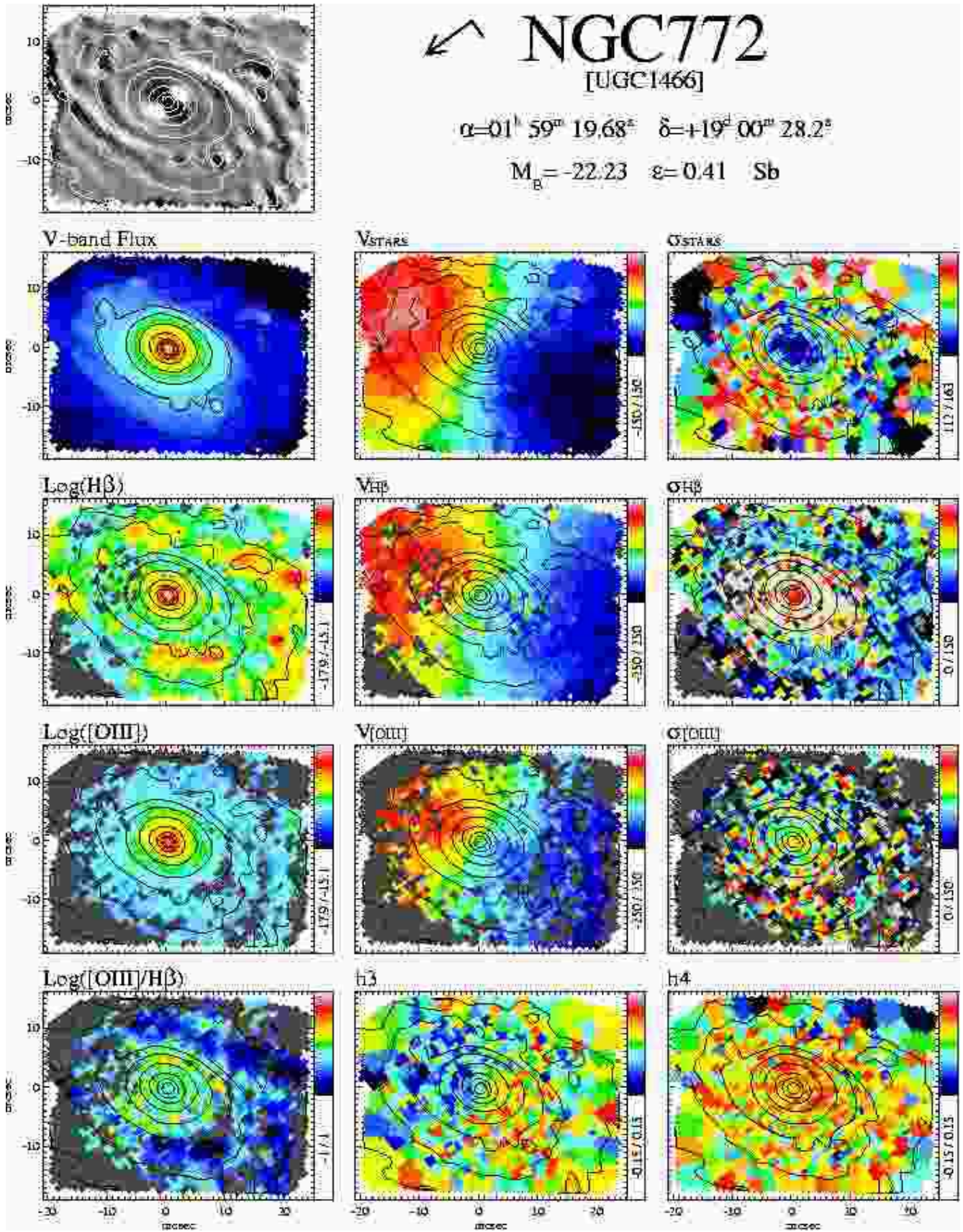}}
\end{center}
\caption{As in Figure \ref{maps1} for NGC772.}\label{maps3}
\end{figure*}
\addtocounter{figure}{-1}
\addtocounter{subfigure}{1}

%ngc864
\clearpage
\begin{figure*}
\begin{center}
{\includegraphics[width=0.99\linewidth]{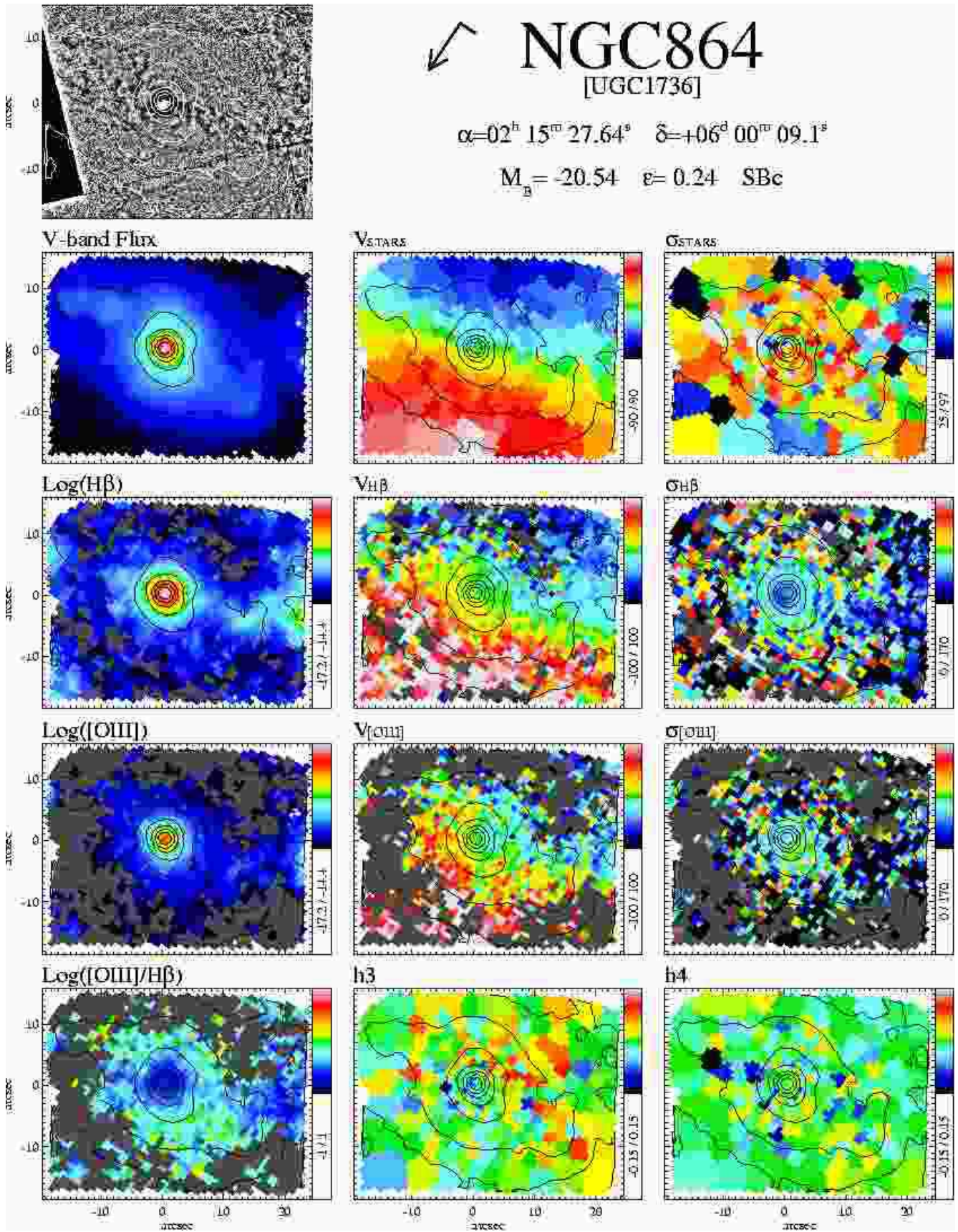}}
\end{center}
\caption{As in Figure \ref{maps1} for NGC864.}\label{maps4}
\end{figure*}
\addtocounter{figure}{-1}
\addtocounter{subfigure}{1}

%ngc1042
\clearpage
\begin{figure*}
\begin{center}
{\includegraphics[width=0.99\linewidth]{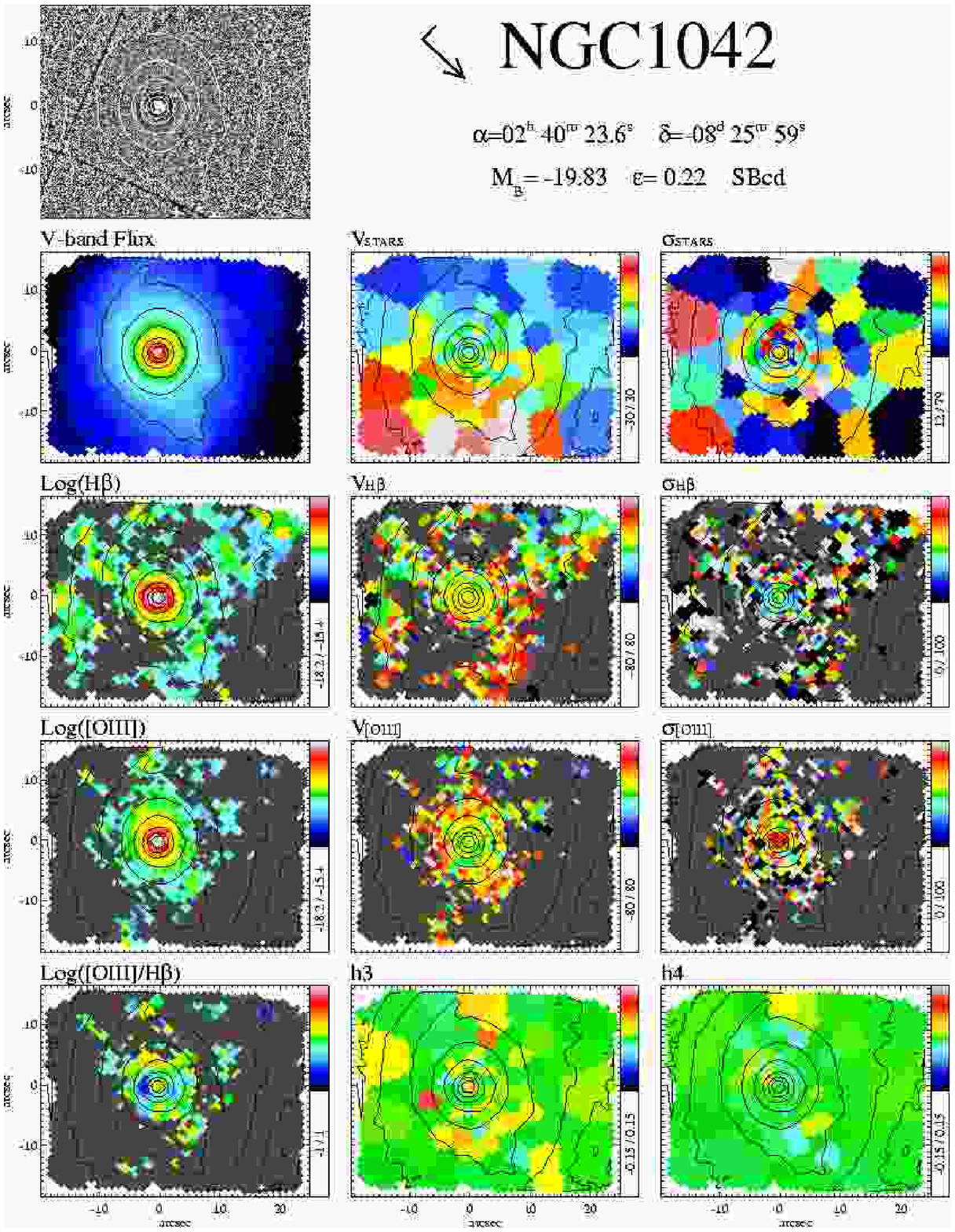}}
\end{center}
\caption{As in Figure \ref{maps1} for NGC1042.}\label{maps5}
\end{figure*}
\addtocounter{figure}{-1}
\addtocounter{subfigure}{1}

%ngc2805
\clearpage
\begin{figure*}
\begin{center}
{\includegraphics[width=0.99\linewidth]{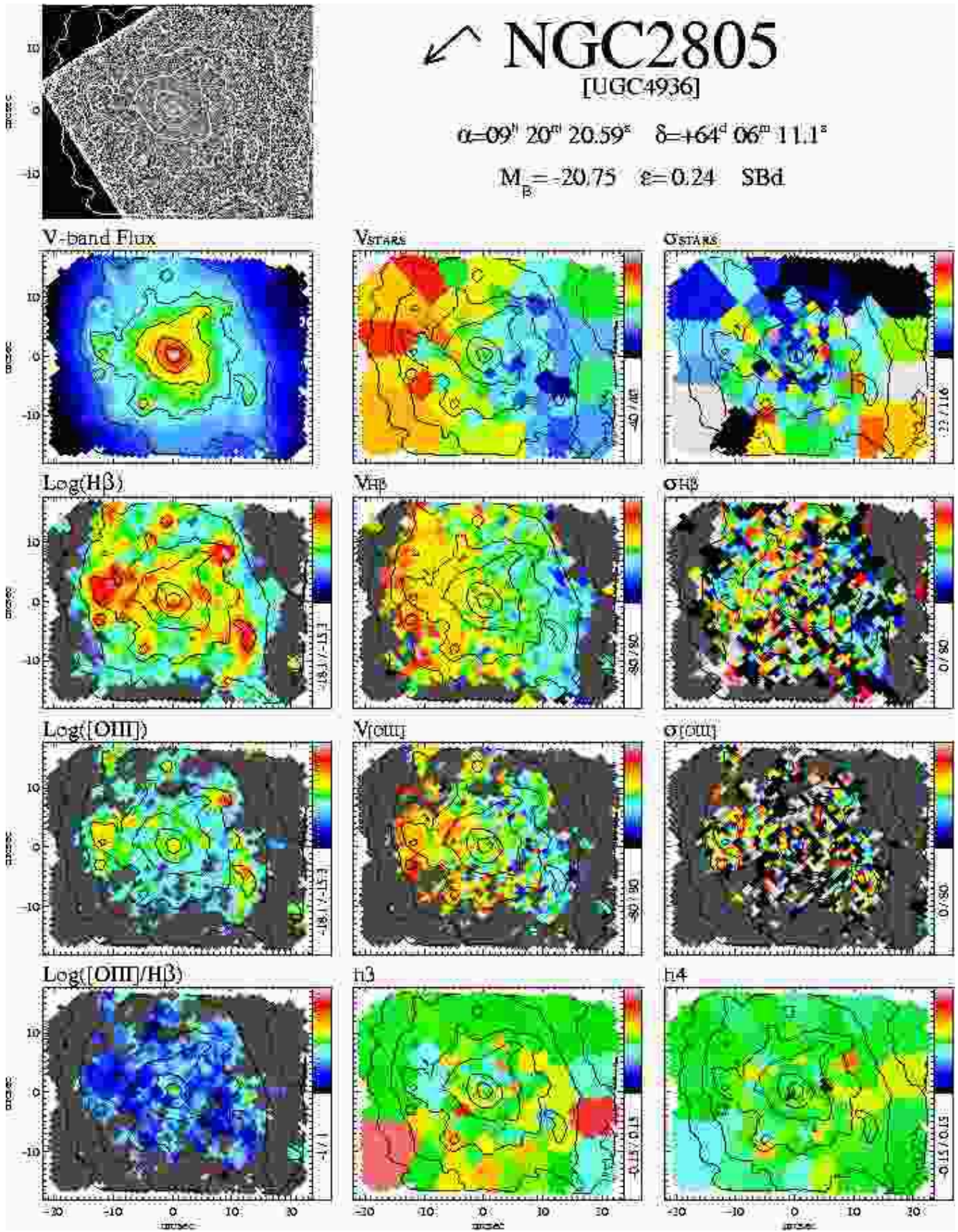}}
\end{center}
\caption{As in Figure \ref{maps1} for NGC2805.}\label{maps6}
\end{figure*}
\addtocounter{figure}{-1}
\addtocounter{subfigure}{1}

%ngc2964
\clearpage
\begin{figure*}
\begin{center}
{\includegraphics[width=0.99\linewidth]{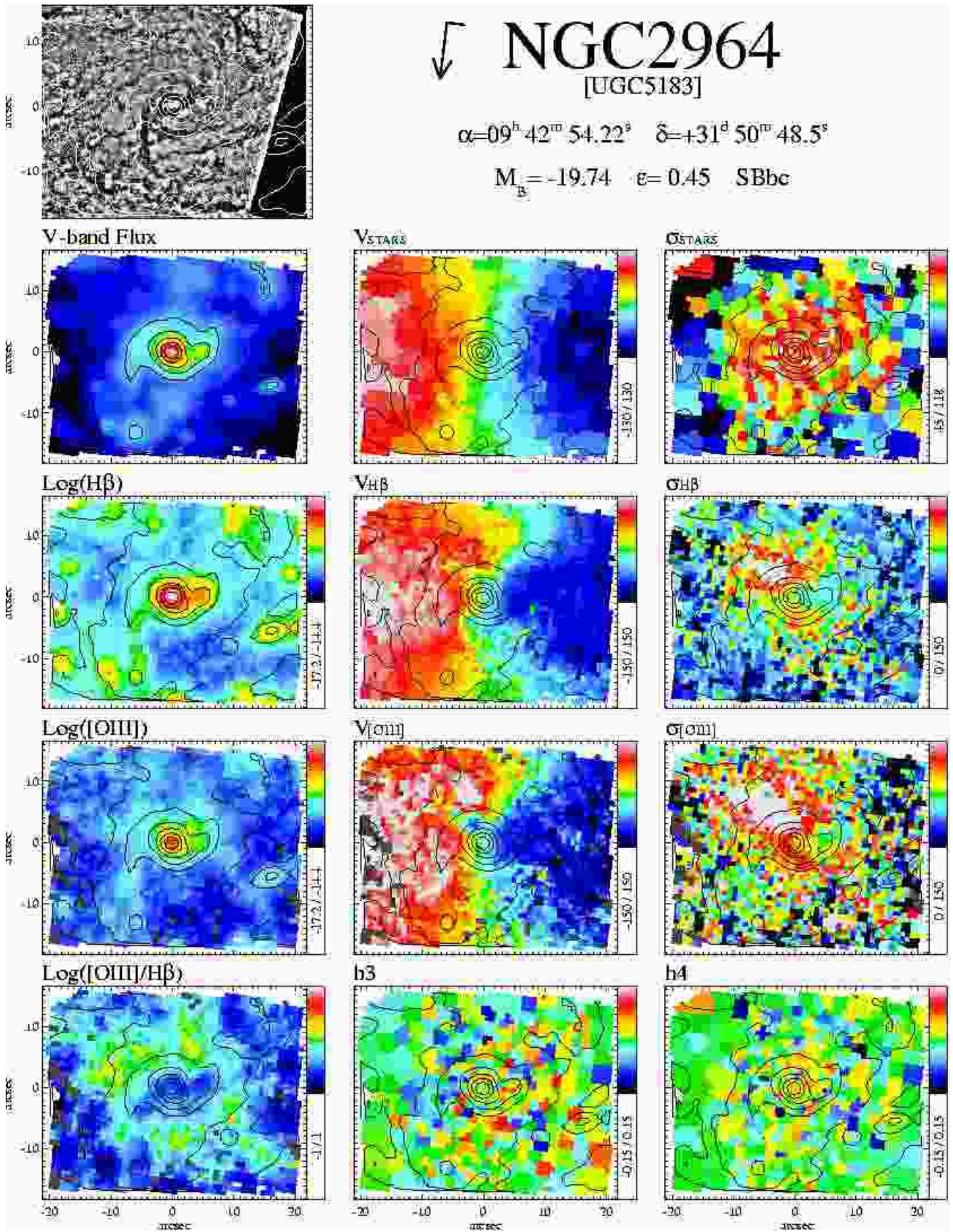}}
\end{center}
\caption{As in Figure \ref{maps1} for NGC2964.}\label{maps7}
\end{figure*}
\addtocounter{figure}{-1}
\addtocounter{subfigure}{1}

%ngc3346
\clearpage
\begin{figure*}
\begin{center}
{\includegraphics[width=0.99\linewidth]{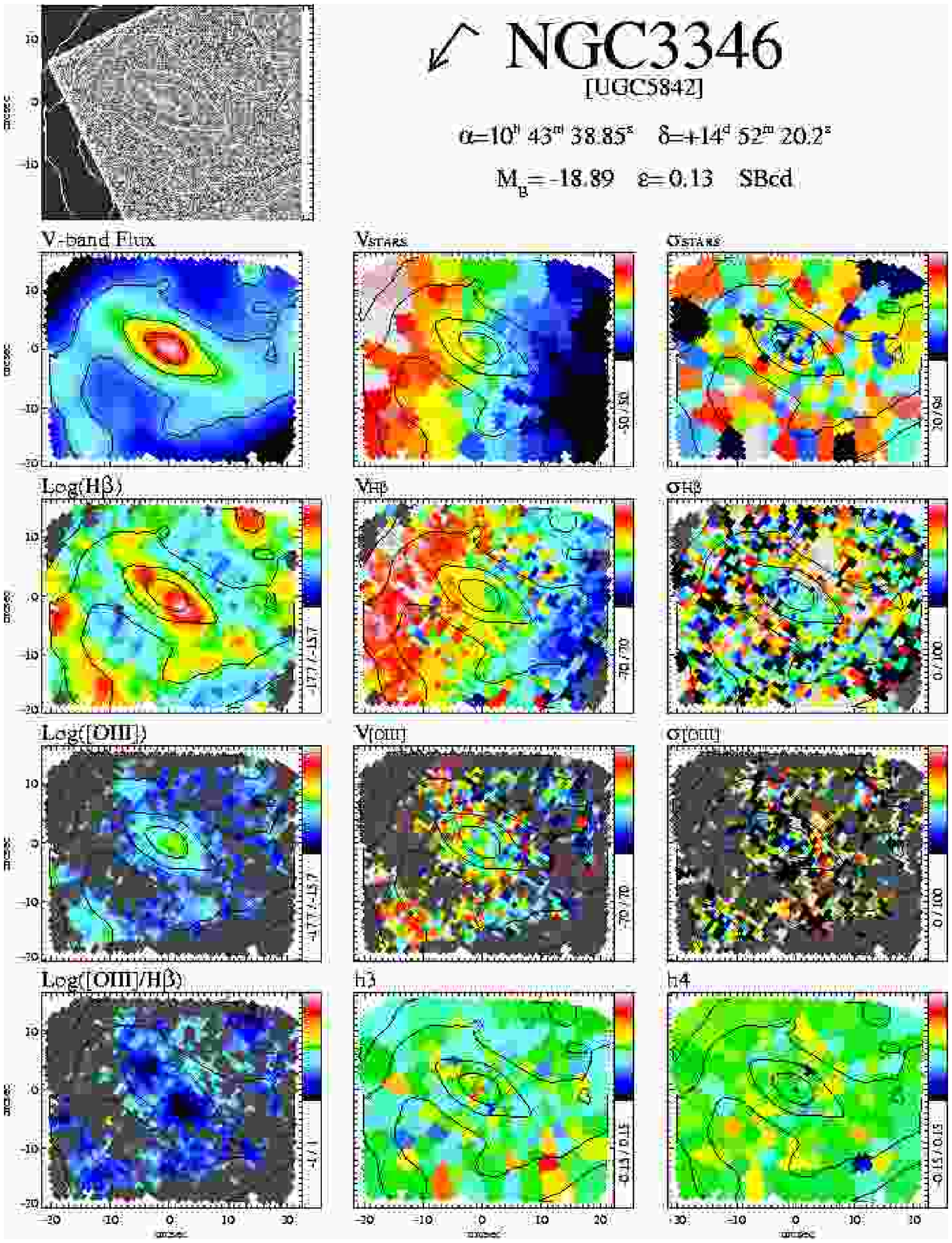}}
\end{center}
\caption{As in Figure \ref{maps1} for NGC3346.}\label{maps8}
\end{figure*}
\addtocounter{figure}{-1}
\addtocounter{subfigure}{1}

%ngc3423
\clearpage
\begin{figure*}
\begin{center}
{\includegraphics[width=0.99\linewidth]{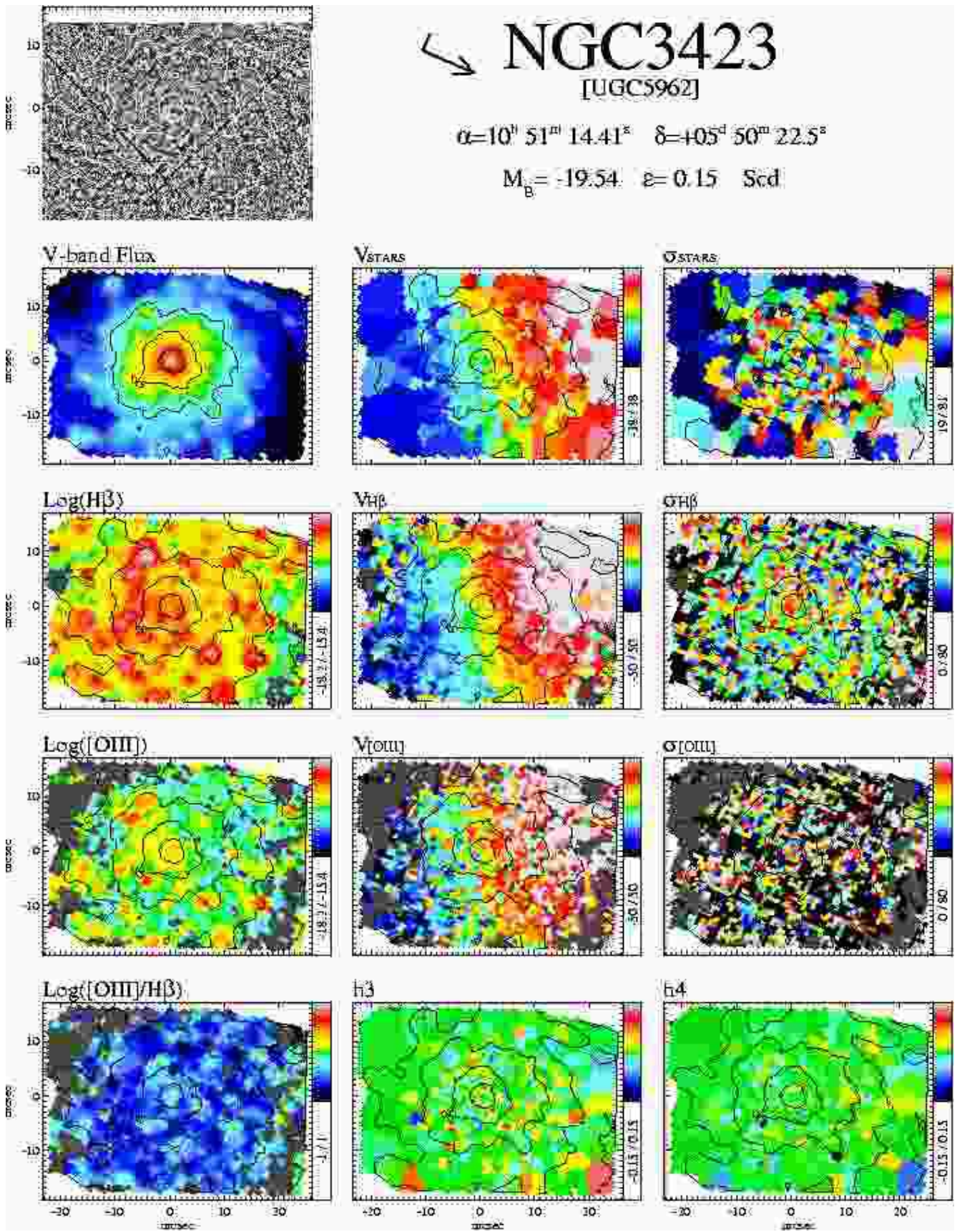}}
\end{center}
\caption{As in Figure \ref{maps1} for NGC3423.}\label{maps9}
\end{figure*}
\addtocounter{figure}{-1}
\addtocounter{subfigure}{1}

%ngc3949
\clearpage
\begin{figure*}
\begin{center}
{\includegraphics[width=0.99\linewidth]{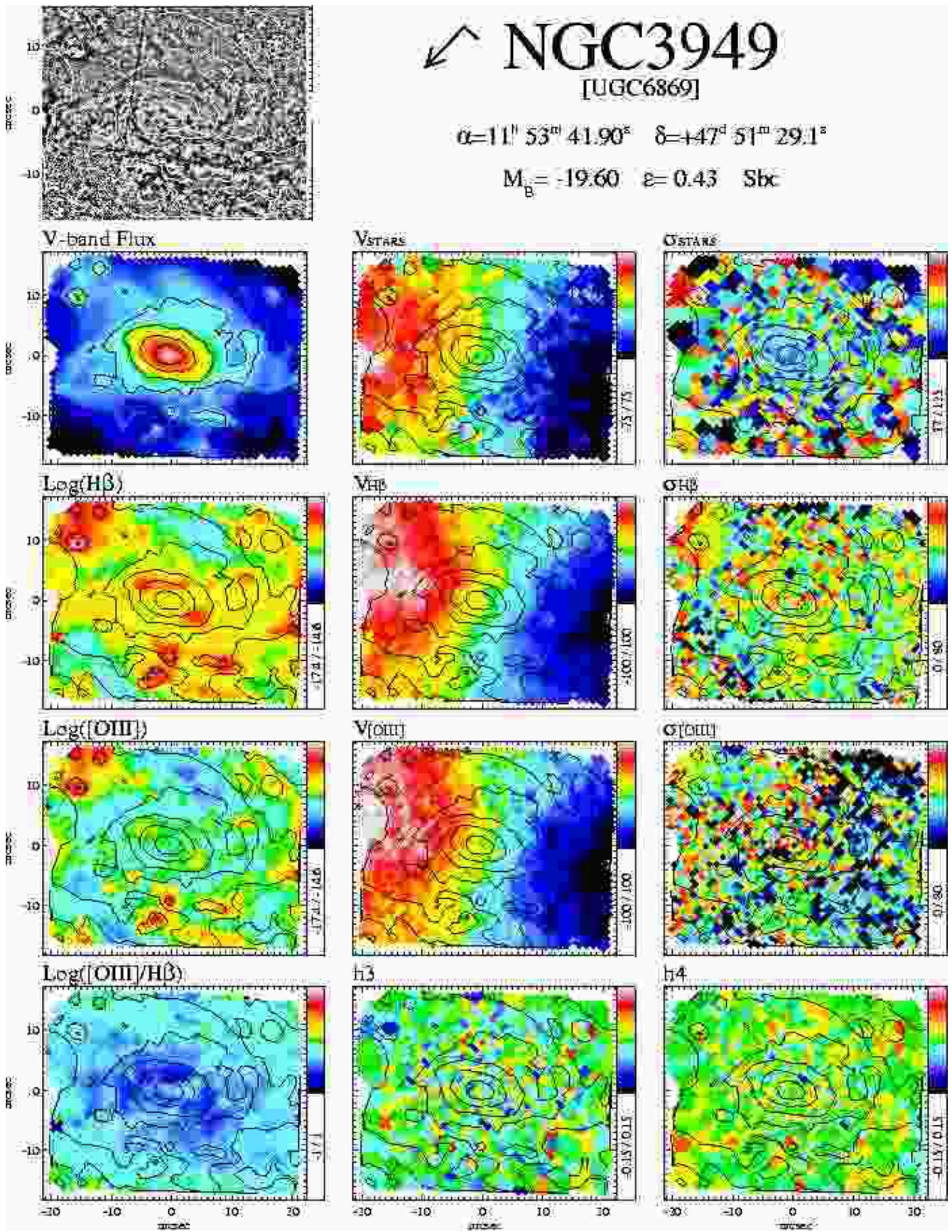}}
\end{center}
\caption{As in Figure \ref{maps1} for NGC3949.}\label{maps10}
\end{figure*}
\addtocounter{figure}{-1}
\addtocounter{subfigure}{1}

%ngc4030
\clearpage
\begin{figure*}
\begin{center}
{\includegraphics[width=0.99\linewidth]{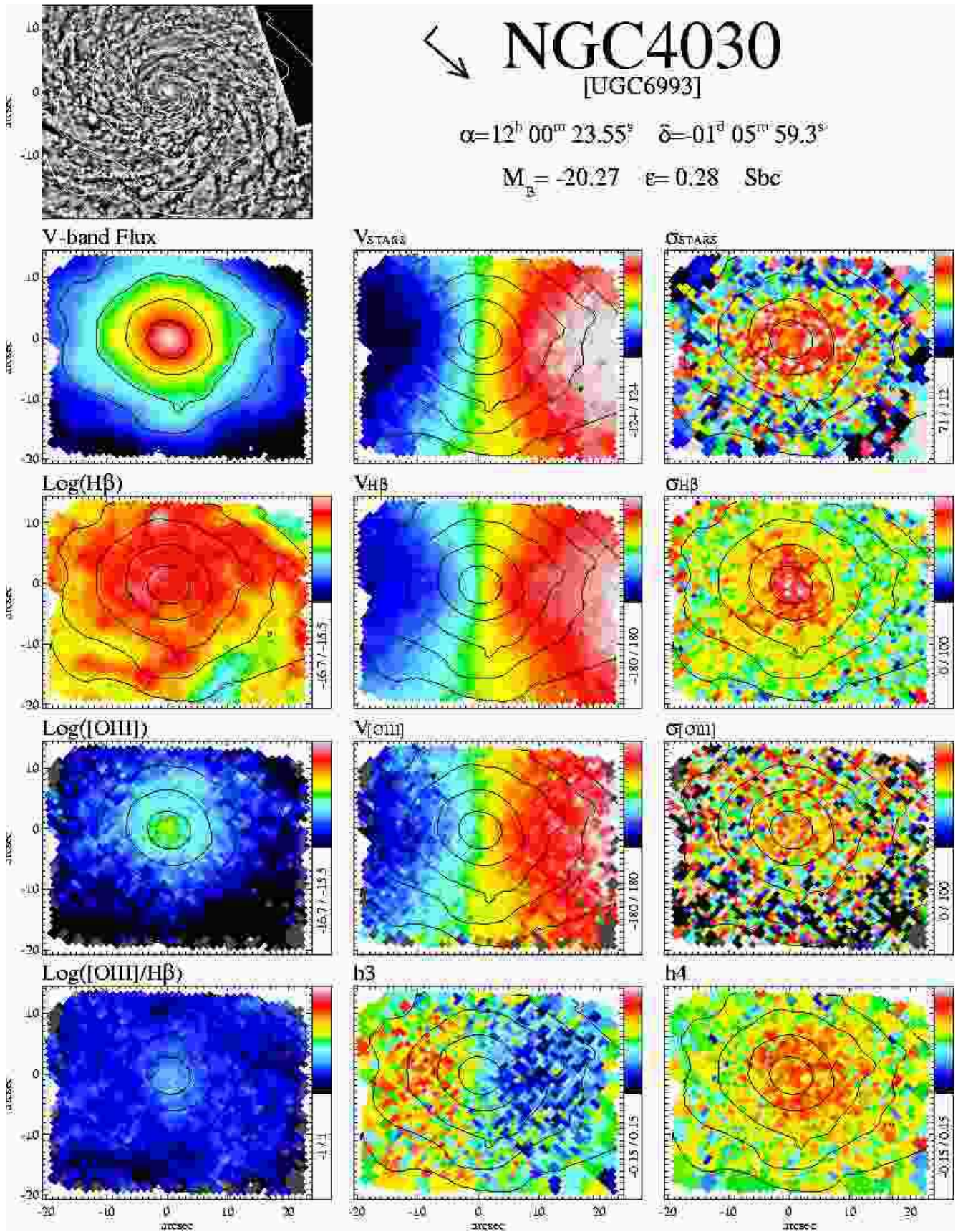}}
\end{center}
\caption{As in Figure \ref{maps1} for NGC4030.}\label{maps11}
\end{figure*}
\addtocounter{figure}{-1}
\addtocounter{subfigure}{1}

%ngc4102
\clearpage
\begin{figure*}
\begin{center}
{\includegraphics[width=0.99\linewidth]{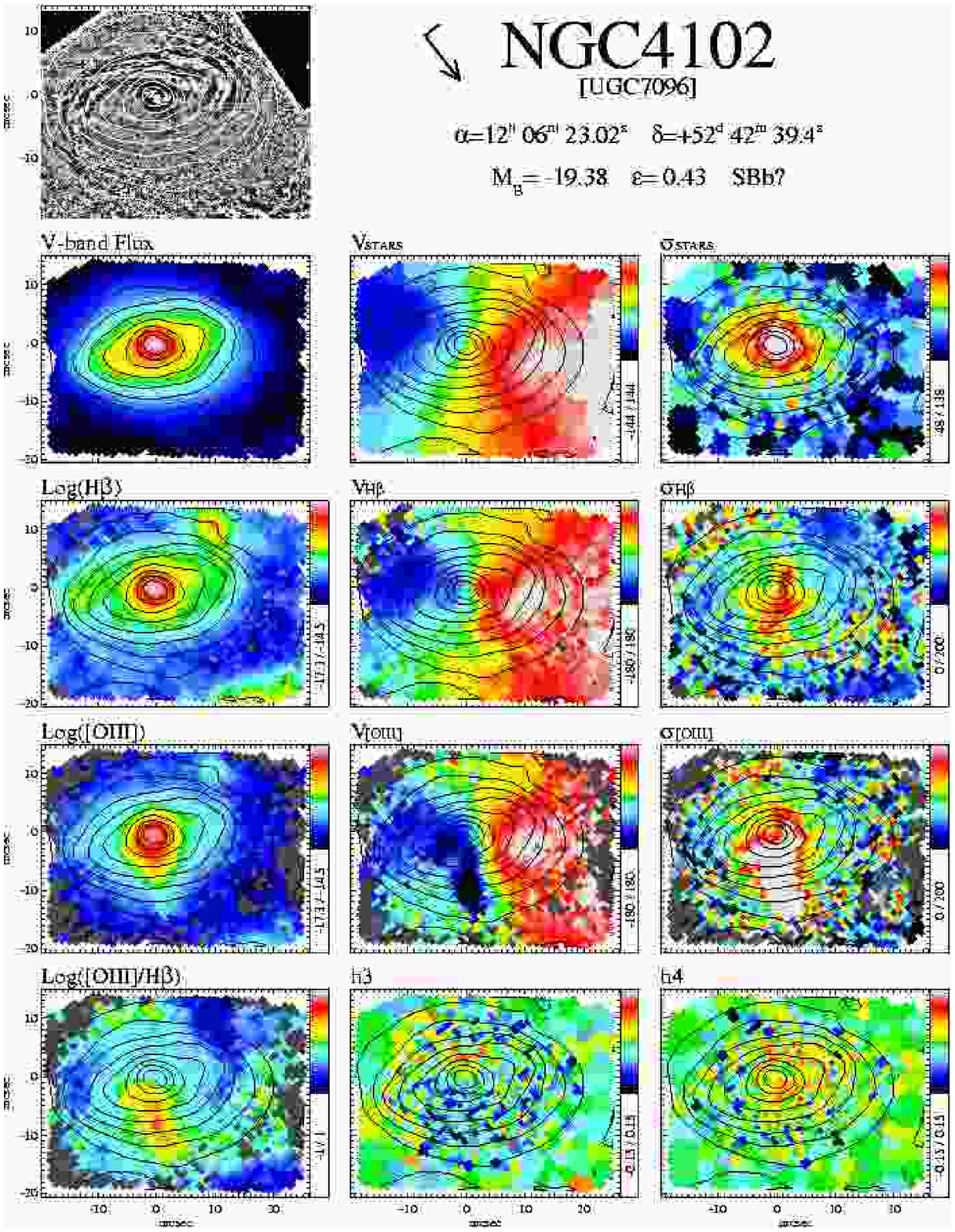}}
\end{center}
\caption{As in Figure \ref{maps1} for NGC4102.}\label{maps12}
\end{figure*}
\addtocounter{figure}{-1}
\addtocounter{subfigure}{1}

%ngc4254
\clearpage
\begin{figure*}
\begin{center}
{\includegraphics[width=0.99\linewidth]{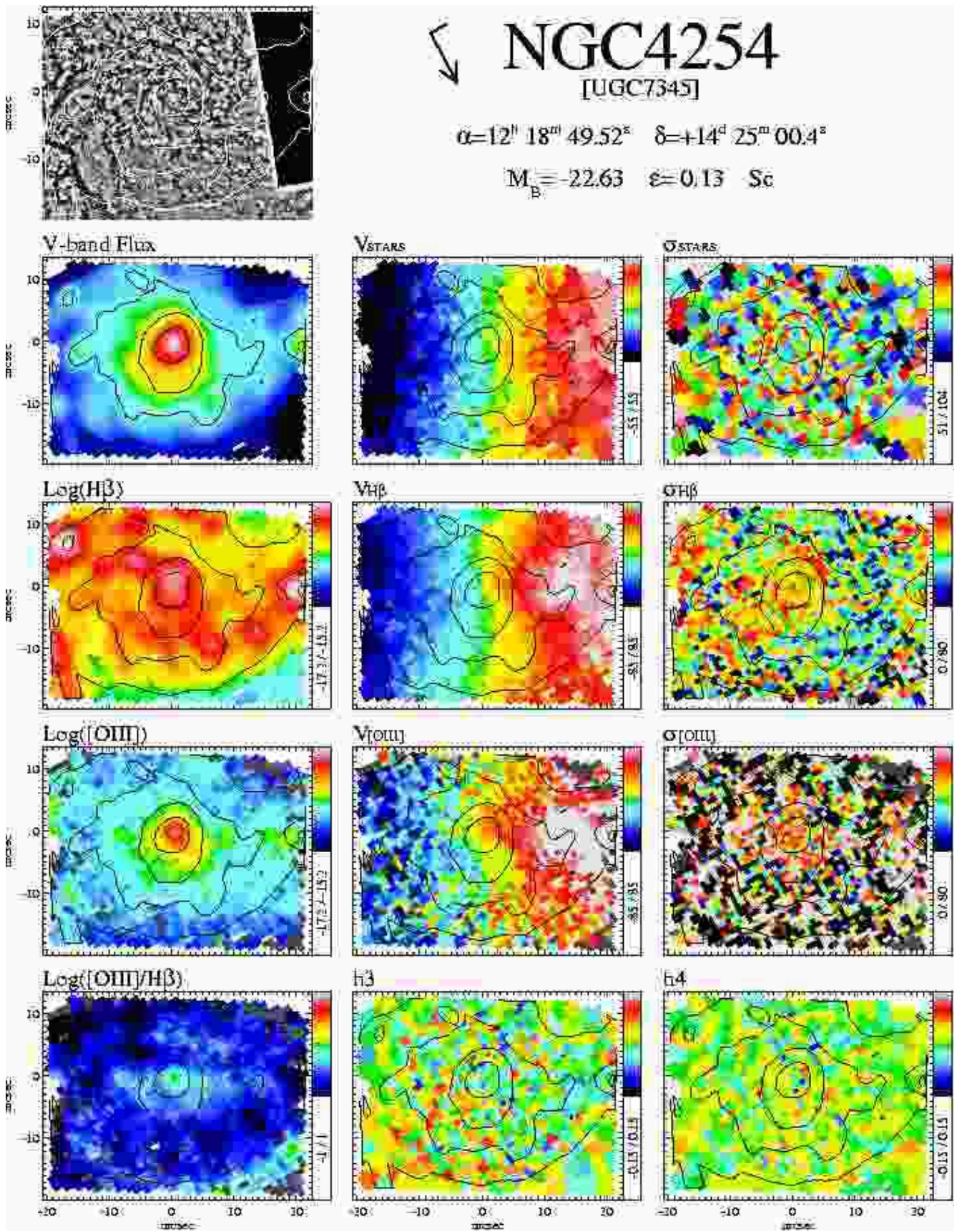}}
\end{center}
\caption{As in Figure \ref{maps1} for NGC4254.}\label{maps13}
\end{figure*}
\addtocounter{figure}{-1}
\addtocounter{subfigure}{1}

%ngc4487
\clearpage
\begin{figure*}
\begin{center}
{\includegraphics[width=0.99\linewidth]{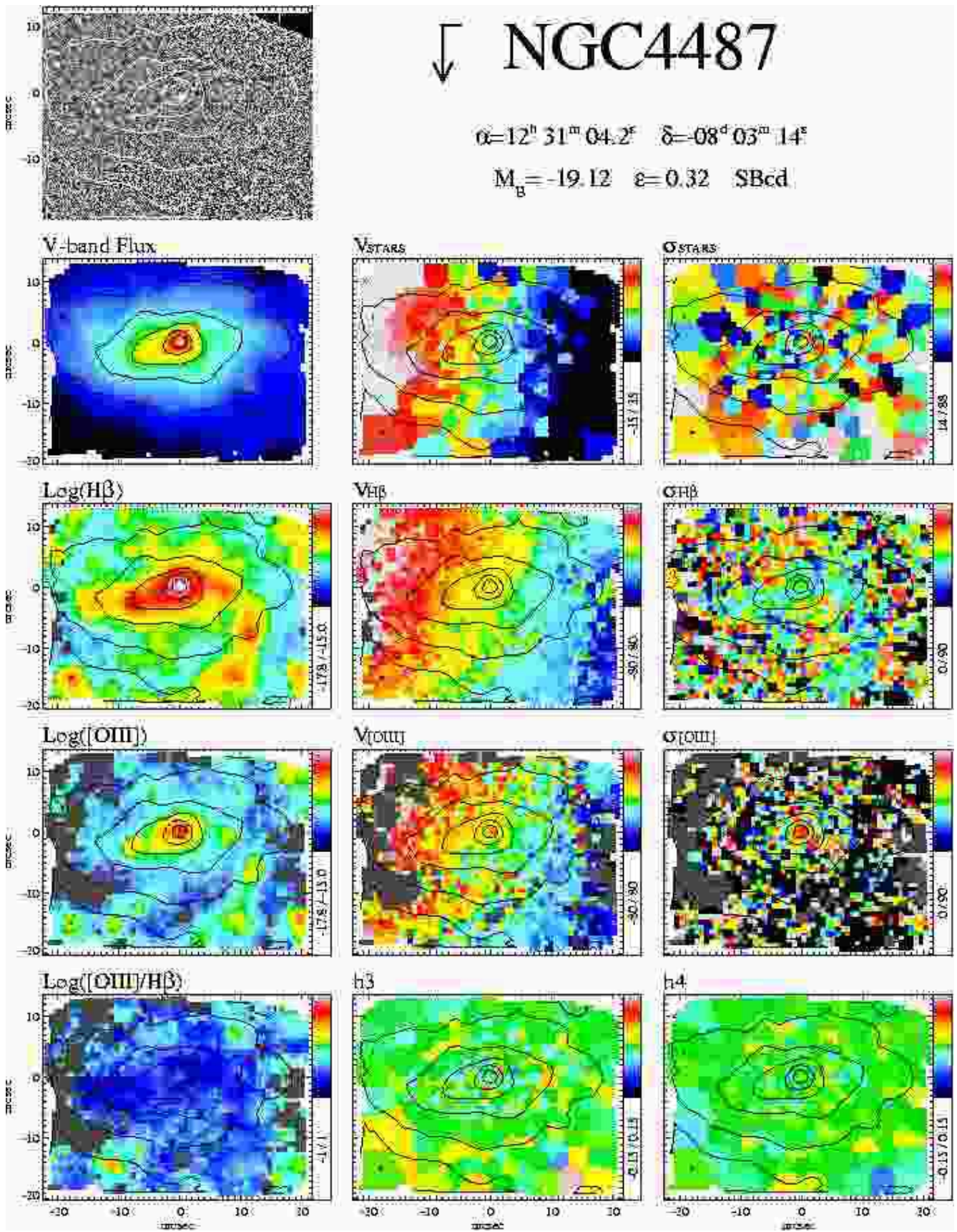}}
\end{center}
\caption{As in Figure \ref{maps1} for NGC4487.}\label{maps14}
\end{figure*}
\addtocounter{figure}{-1}
\addtocounter{subfigure}{1}

%ngc4775
\clearpage
\begin{figure*}
\begin{center}
{\includegraphics[width=0.99\linewidth]{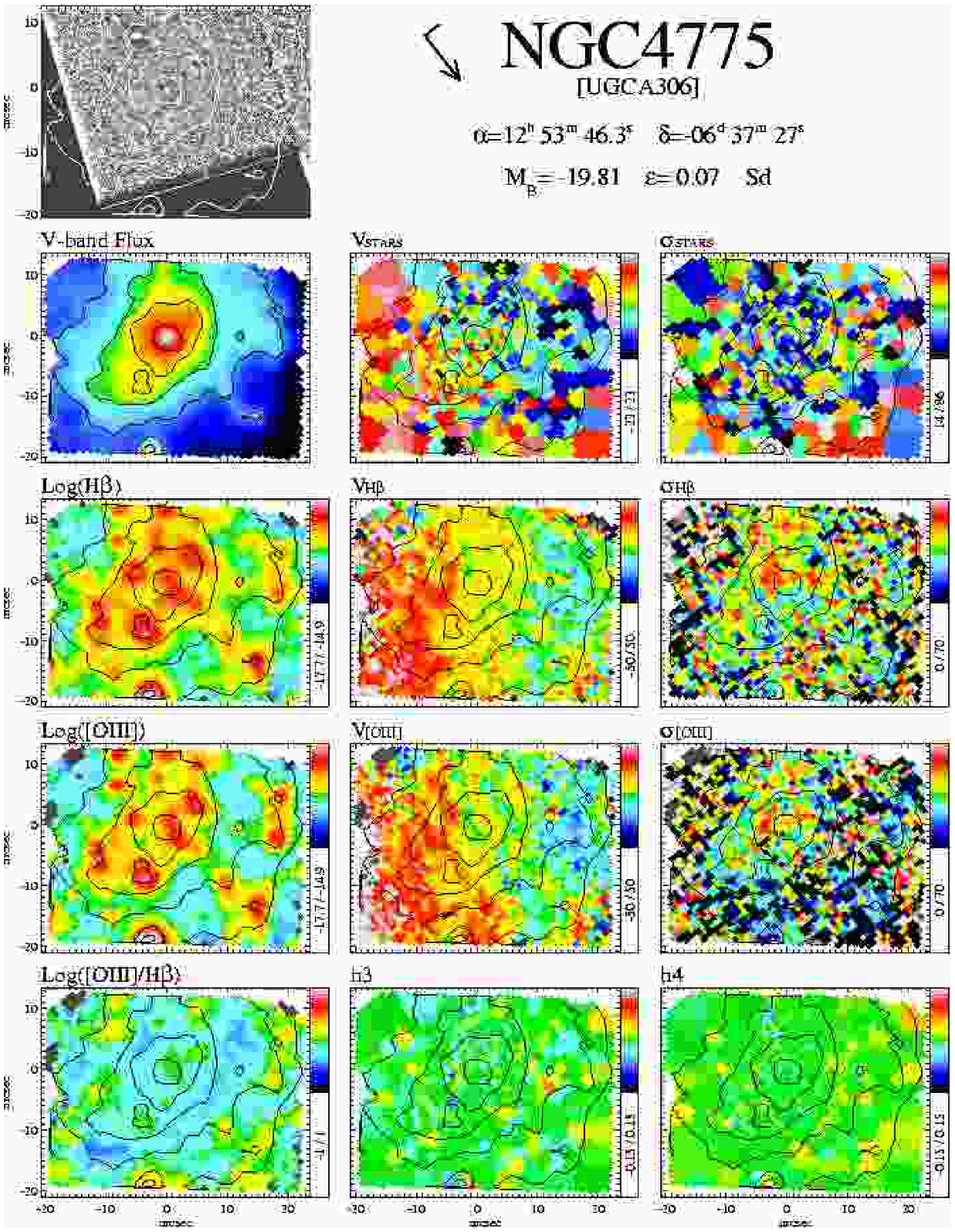}}
\end{center}
\caption{As in Figure \ref{maps1} for NGC4775.}\label{maps15}
\end{figure*}
\addtocounter{figure}{-1}
\addtocounter{subfigure}{1}

%ngc5585
\clearpage
\begin{figure*}
\begin{center}
{\includegraphics[width=0.99\linewidth]{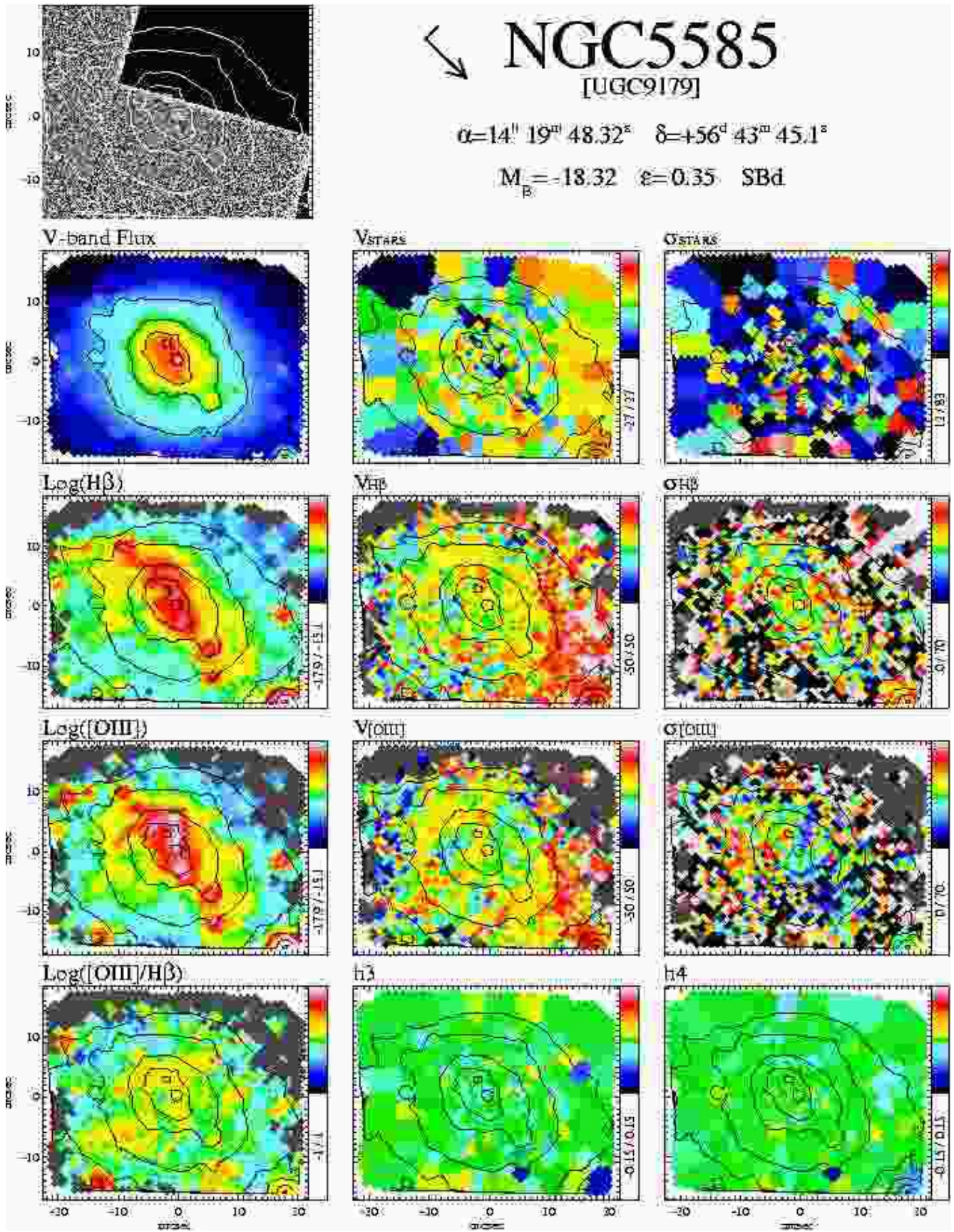}}
\end{center}
\caption{As in Figure \ref{maps1} for NGC5585.}\label{maps16}
\end{figure*}
\addtocounter{figure}{-1}
\addtocounter{subfigure}{1}

%ngc5668
\clearpage
\begin{figure*}
\begin{center}
{{\includegraphics[width=0.99\linewidth]{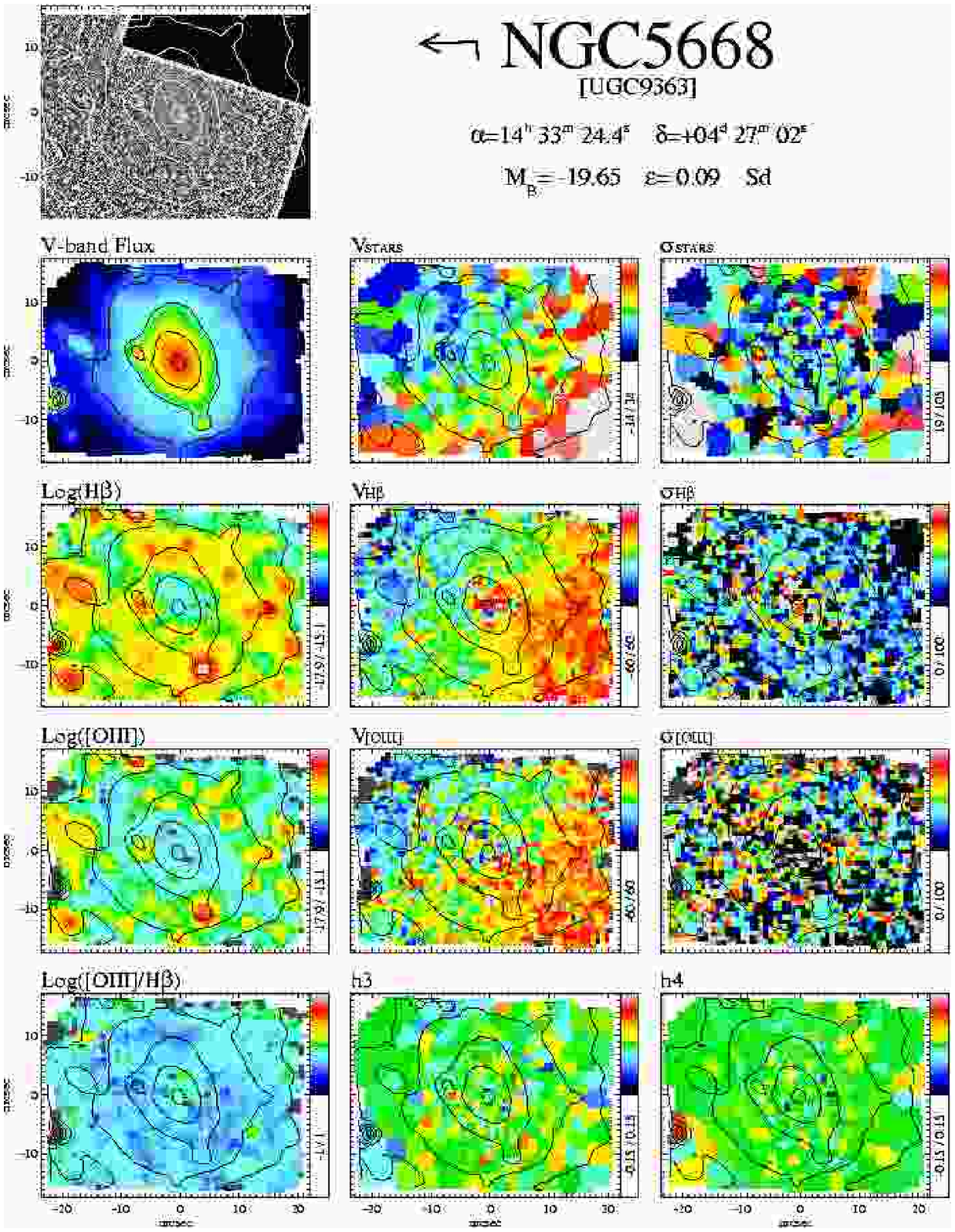}}}
\end{center}
\caption{As in Figure \ref{maps1} for NGC5668.}\label{maps17}
\end{figure*}
\addtocounter{figure}{-1}
\addtocounter{subfigure}{1}

%ngc5678
\clearpage
\begin{figure*}
\begin{center}
{\includegraphics[width=0.99\linewidth]{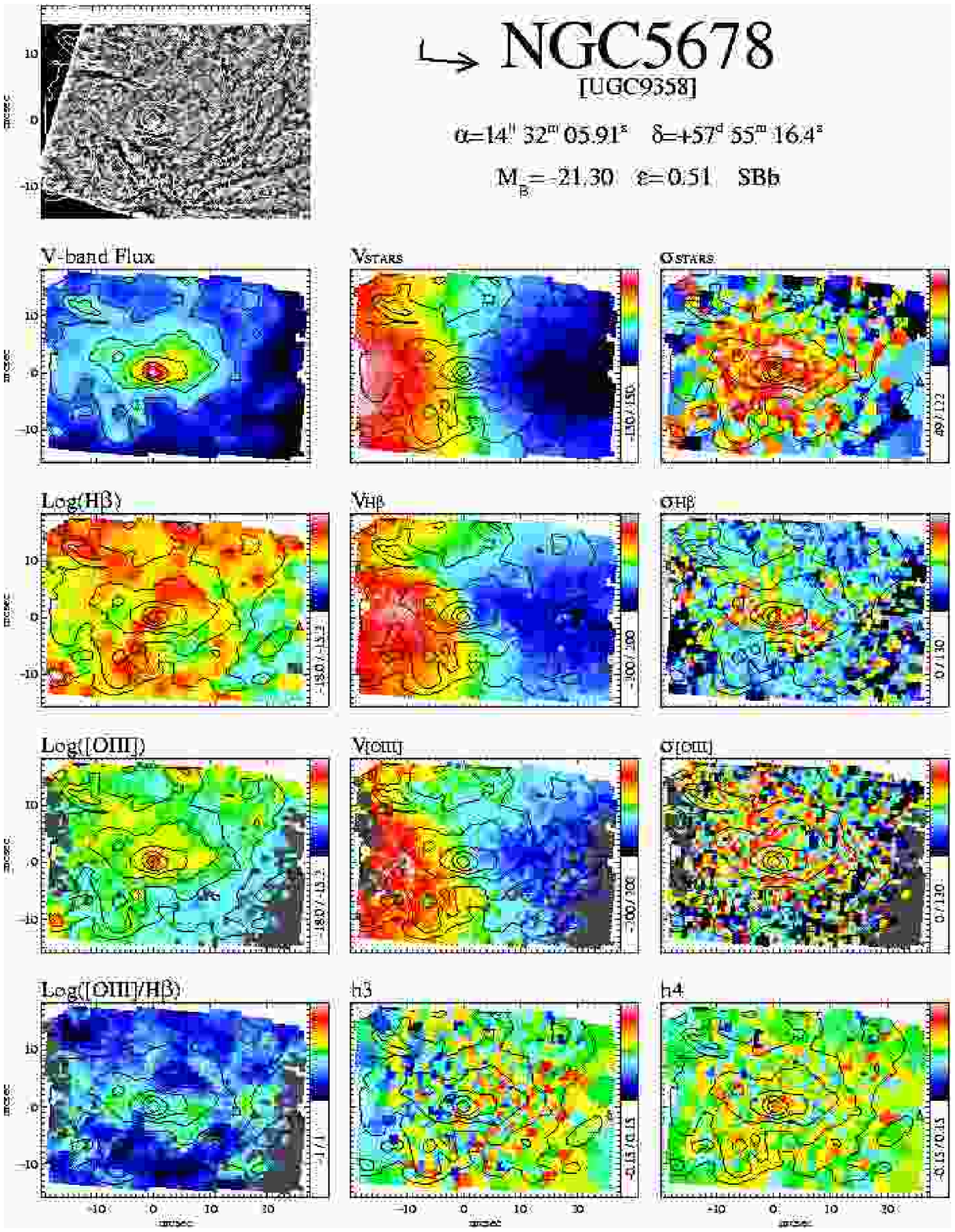}}
\end{center}
\caption{As in Figure \ref{maps1} for NGC5678.}\label{maps18}
\end{figure*}
\clearpage

\subsection{Stellar kinematics}\label{stkinobs}
All the objects show rotation, as expected. An interesting feature is the quite frequent central decline in $\sigma$ (NGC628, NGC772,
NGC2805, NGC3346, NGC3949, NGC5668). We measured also the 
 higher moments $h_{3}$ and $h_{4}$, but unfortunately our spectra 
 contain information only for the most early-type galaxies in our sample 
 (NGC488, NGC772, NGC4030, NGC4102) which are generally the objects with the highest $(S/N)_\star$, where penalization does
 not play an important role (see Section \ref{methodstar} and \citealt{ppxf}). 
 Some objects display misaligned photometric and kinematical axes (NGC864, NGC3346, NGC4487), a possible indication of non-axisymmetric structures such as bars. 
In other cases, the situation is less clear, due to the presence of twists in the rotation axis (NGC772, NGC2964, NGC3949, NGC4254, NGC5678). In many galaxies our measurements 
indicate very low velocity dispersions, as seen also on the basis of the central values reported in Table \ref{properties} and in Figure \ref{sampleprop}. 

\subsubsection{Radial behaviour of the stellar velocity dispersion }
We addressed the radial behaviour of the stellar velocity dispersion $\sigma$ by measuring the radial $\sigma$ profiles and their slope and
correlating this with
the morphological type. For each of our galaxies, we computed a $\sigma$ profile by averaging the stellar velocity dispersion map on elliptical annuli orientated as the 
galaxy isophotes; for some strongly barred galaxies, the chosen orientation does not coincide with the PA quoted in Table \ref{properties}, which refers to the outermost
isophotes. The ellipticity of the concentric ellipses is instead taken in all cases from Table \ref{properties}. The distance between consecutive annuli is $0\farcs8$ along the minor axis, corresponding to the pixel size of the unbinned {\tt SAURON} cubes. 
In order to take into account the galaxy size, we rescaled the radial coordinate by dividing it by the disc scale length. The values for the disc scale length $h_{r}$ come from one-dimensional 
fitting of the photometric profiles that we extracted from space- and ground- based images via isophotal analysis. The details of this photometric analysis will be given 
in a future paper. We then estimated the $\sigma$ gradient across 
the field by fitting a straight line to the data points by means of a least-squares algorithm. Figure \ref{sigma_gradients} illustrates the results. For each galaxy the computed $\sigma$ profile is plotted against  
the scale-free radius $r/h_{r}$; the solid lines overplotted are the best-fitting straight lines, the slope of which is indicated in a corner of each panel. One
can see that for some galaxies a 
straight line is clearly not an optimal description of the data, at least not over the whole radial range considered, but it
serves as an indication of a global trend. 
The dotted line drawn on each plot marks the $r=1\farcs 2$ line, which represents the edge of the squared aperture within which the central $\sigma$ values reported in Table \ref{properties} were
computed. In the following Figure \ref{sigma_type} we plot the slope of these scale-free fits against the morphological type of the
galaxy. A weak global trend 
can be recognized: the slope tends to increase with later types, indicating that it is more probable for later-type galaxies to have a central
region colder than the surroundings, rather than a hotter one. In the Figure, we
labelled with (B) the galaxies classified as barred (B and AB in Table \ref{properties}), but we do not detect any
significant correlation with bar classification.\\
\indent The velocity dispersion profile is determined
by the mass distribution of the galaxy, the anisotropy of the velocity distribution and the viewing angle. Galaxies with more concentrated light 
distribution will have larger central peaks in their $\sigma$ profiles. Elliptical galaxies generally have outwards decreasing $\sigma$ profiles (see
for example \citealt{onofrio}), 
and this is also the case for many early-type spirals \citep{jesus}, as a result of a centrally concentrated bulge. Here we see that for the
later type spiral galaxies there is no sign anymore in the $\sigma$ profile of this central mass concentration, which is indeed
expected for galaxies with lower bulge/disc ratios: for later types bulges are 
smaller and have lower surface brightness (see for example \citealt{jesus02}). In some cases, however, we see that $\sigma$ is 
increasing outward, which is a sign of cooler central mass concentration. \citet{eric01}, \citet{marque}, \citet{shapiro} and Paper VII have shown that these central $\sigma$-drops are quite common, and are in
general associated with disc-like structures. They are most likely due to stellar discs formed from gas that has been accreted towards the
centre. This is in agreement with simulations by \citet{woz}. Our measurements show that also galaxies as late as Sd present velocity dispersion drops.
The fact that the velocity dispersion profiles are flat or rising does not necessarily mean that these galaxies have no bulge - by detailed
modelling in a future paper we will be able to answer this question.
\renewcommand{\thefigure}{\arabic{figure}}
\begin{figure*}
{\includegraphics[width=0.99\linewidth]{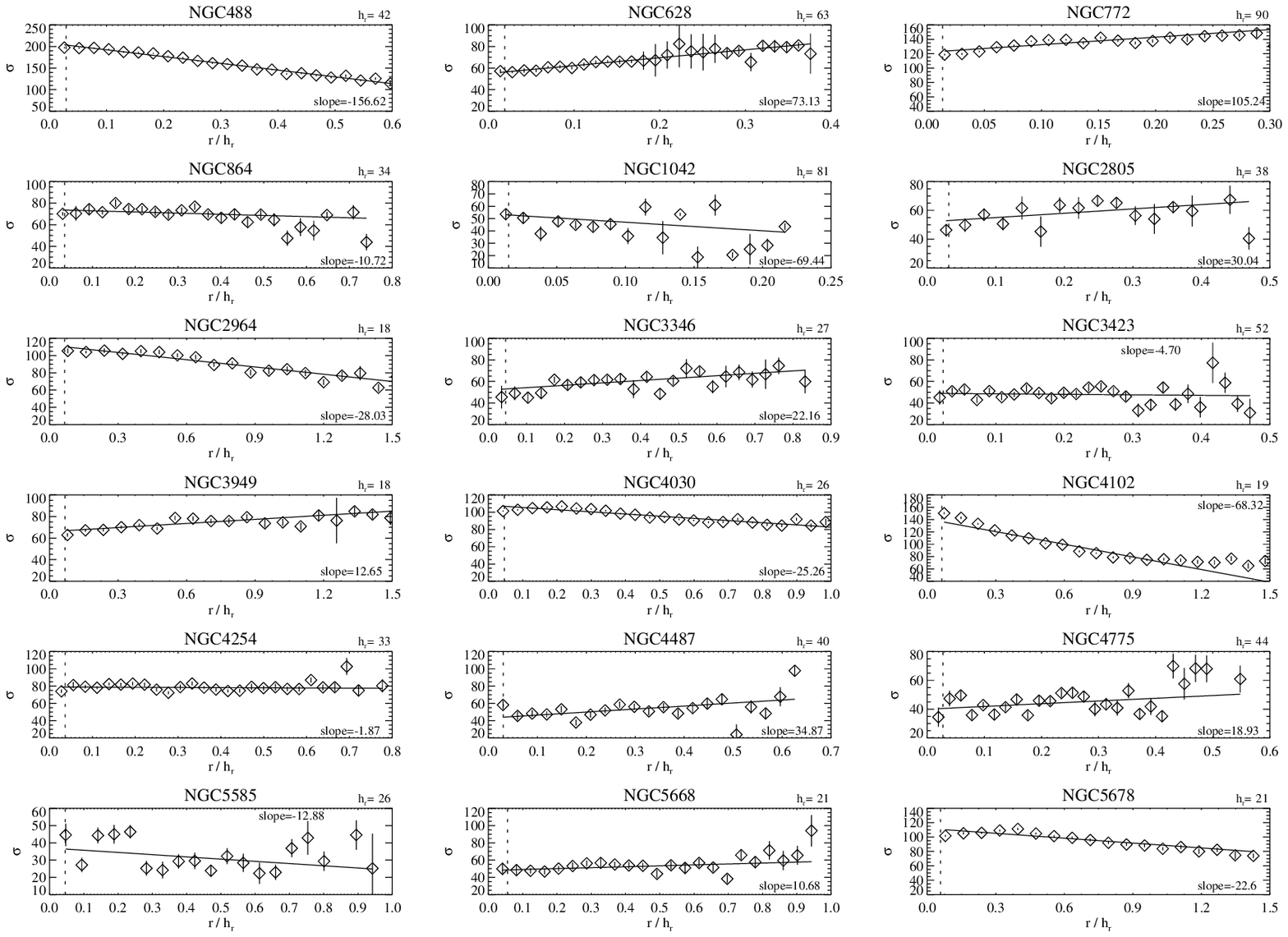}}
\caption{The plots present galaxy by galaxy the radial profile of stellar velocity dispersion (in km s$^{-1}$), calculated on elliptical annuli as explained in the text; 
on the horizontal axis we plot a scale-free radius, obtained dividing our radial coordinate (semi-major axis of the ellipses in arcsec) 
by the disc scale-length $h_{r}$ (also expressed in arcsec and written above each plot).
A straight-line is fitted to the velocity dispersion profiles; the solid lines overplotted are the best-fitting straight lines, the slope of which is
recorded in a corner of each panel; the dotted vertical line indicates in each panel the 
$r=1\farcs2$ line within which the central values 
of Table \ref{properties} were calculated.}
\label{sigma_gradients}
\end{figure*}
\begin{figure*}
{\includegraphics[width=0.69\linewidth]{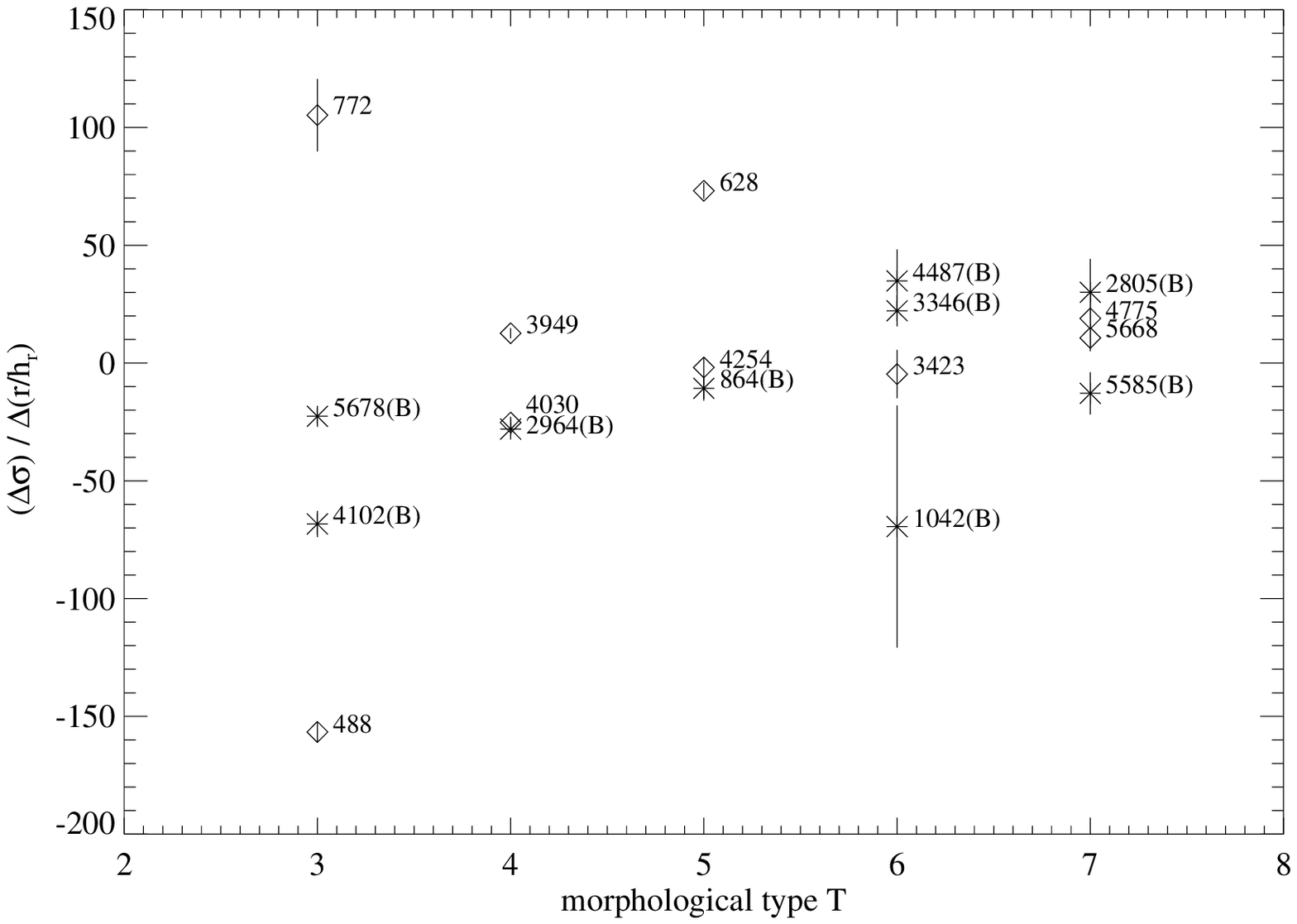}}
\caption{The fitted slope $\Delta\sigma/\Delta (r/h_{r})$ 
of the velocity dispersion profiles is plotted here as a function of morphological type; the values for this slope are those indicated in the
panels in Figure \ref{sigma_gradients}. The galaxies classified as barred are here labelled with (B) and represented by asterisks; 
the NGC galaxy identifiers are indicated close to the corresponding symbol.}
\label{sigma_type}
\end{figure*}

\subsubsection{Qualitative comparison with HST photometry}
It is not straightforward to connect our kinematic results with photometry. In the last decade analysis based on HST imaging
(\citealt{marcella97}, 1998, 2002, \citealt{reynier}, \citealt{boker}, \citealt{laine}) revealed that late-type galaxies do not necessarily possess
classical bulges\footnote{We note here that Carollo (1997, 1998, 2002) refers to a definition of classical bulge as a centrally concentrated stellar distribution with an
amorphous, smooth appearance. This implicitly excludes gas, dust, and continuing recent star formation \citep{wyse}.}; they can instead host small-scale structures such as nuclear 
bars, nuclear star clusters, point-like sources. We cross-correlated the above mentioned samples with our own and tried to look for kinematical signatures of the 
photometric features that they detected in the images. In Table \ref{photometry} we tabulate the results of this comparison, based on a review of the relevant literature, on 
the appearance of our maps (Figures \ref{maps1}-\ref{maps18}, Section \ref{allthemaps}) and of the radial behaviour of $\sigma$ as well (Figure \ref{sigma_gradients}).\\ 
\begin{table}
\begin{center}
 \begin{tabular}{@{}l l l l}
  \hline\hline
 \small{NGC} & \small{HST} &\small{ref} & \small{{\tt SAURON}}\\
 \hline
 \small{488}&\small{r$^{1/4}$b}&\small{C97} & \small{hot central region}\\
 \small{1042}&\small{nsc}&\small{B02}& \small{low $\sigma$ over the field}\\
 \small{2805}&\small{nsc}&\small{B02}& \small{cold central region} \\
\small{2964}&\small{sf+nuc}&\small{C97, C02}& \small{$\sigma$ enhanced in the centre}\\
\small{3346}&\small{nsc}&\small{B02} & \small{cold central region} \\  
\small{3423}&\small{nsc}&\small{B02}& \small{low $\sigma$ over the field} \\ 
\small{3949}&\small{nuc}&\small{C02}& \small{cold central region} \\
\small{4030}&\small{irb+expb}&\small{C98}& \small{hot central region}\\
\small{4102}&\small{irb}&\small{C97}& \small{hot central region}\\
\small{4487}&\small{nsc}&\small{B02} & \small{cold central region}\\
\small{4775}&\small{nsc}&\small{B02} &\small{low $\sigma$ over the field}\\
\small{5585}&\small{nsc}&\small{B02} &\small{low $\sigma$ over the field}\\
\small{5668}&\small{nsc}&\small{B02} &\small{cold inner region}\\
\small{5678}&\small{nuc}&\small{C02}& \small{hot central region} \\
 \hline
 \end{tabular}\\
 \end{center}
 \caption{Comparison with HST photometry. References in the table: C97 $=$ \citealt{marcella97}, C98 $=$ \citealt{marcella98b}, C02 $=$ \citealt{marcella02}, B02 $=$ \citealt{boker}; abbreviations:
 r$^{1/4}$b$ =$ de Vaucouleurs bulge, irb $=$ irregular bulge (when a galaxy possess a central dense component but does not have a featureless appearance), expb $=$ exponential bulge, 
 nuc $=$ nuclear compact source, nsc $=$ nuclear star cluster, sf $=$ nuclear star forming
 region; for a more accurate description of the photometric features, we refer the reader to the quoted papers.}
\label{photometry}
\end{table}
\indent According to \citet{boker}, NGC1042, NGC2805, NGC3346, NGC3423, NGC4487, NGC4775, NGC5585 and NGC5668 host a nuclear star cluster and do not have a stellar 
bulge; our data show that the central regions of these
objects are indeed cold. For any speculation on nuclear star clusters, caution is needed, since the 
nuclear star clusters detected with HST imaging have a spatial scale usually much smaller than the {\tt SAURON} resolution.   
There are also cases where our observations nicely support 
the photometrical results: NGC488 has a classical r$^{1/4}$ bulge \citep{marcella97}, and our maps show a hot, extended central region; NGC4030 has also a 
large stellar bulge - although classified as  ``irregular'' and ``exponential'' \citep{marcella98b} - and our data show a hot nuclear region. 
To conclude, the examples show that there is a correspondence between the photometric lack of a stellar bulge and the kinematical measurement that the inner
regions are cold. At the moment this is a qualitative statement that requires further investigation. As previously stated, in a future paper we will study deeper the relation between the kinematical 
structures and the shape of the photometrical profiles.

\subsection{Gas kinematics}\label{gaskinobs}
\begin{table}
\begin{center}
 \begin{tabular}{l c c l c c}
  \hline\hline
 NGC & $\log$F\tiny{H$\beta$}  & $\log$F\tiny{[OIII]} & NGC & $\log$F\tiny{H$\beta$}  & $\log$F\tiny{[OIII]}\\
 \hline
 488 & -13.72&-13.79 &3949& -12.32&-12.59\\
 628 & -13.34&-14.12 &4030 & -12.51&-13.18\\
 772 & -13.18&-13.59 &4102 & -12.69&-12.85\\
 864 & -13.03&-13.53 &4254& -12.48&-13.13\\
 1042 & -13.84&-13.92 &4487& -12.91&-13.40\\
 2805& -13.32& -13.84 &4775& -12.64&-12.70\\
 2964 & -12.60& -13.03 &5585& -12.98&-12.89\\
 3346 & -13.47&-14.14 &5668& -12.89&-13.15\\
 3423 & -12.93&-13.45 &5678& -12.80&-13.33\\
 \hline
 \end{tabular}\\
 \end{center}
 \caption{The table reports galaxy by galaxy the decimal logarithm of the 
 global fluxes in the H$\beta$ and [OIII] emission-lines, in erg s$^{-1}$ cm$^{-2}$.}
\label{gasflux}
\end{table}
In our data, the gas rotates usually faster than the stars and around the same axis; the gas velocity fields are much more complicated than the stellar ones, 
with zero-velocity lines that are often wiggling, inclined or S-shaped (NGC772, NGC2964, NGC3346, NGC3949, NGC4102, NGC4254, NGC5678). When comparing the results in the two 
diagnostic lines (H$\beta$ and [OIII]), we notice that in many cases the H$\beta$ distribution is spatially more extended than the [OIII] one (NGC488, NGC628, NGC864, 
NGC1042, NGC2805, NGC3346) and that the flux in H$\beta$  is in most cases higher than in [OIII], although there can be specific regions where 
the situation is opposite, with less emission in H$\beta$ rather than in [OIII], which is the case for NGC4102 and NGC5585. In Table \ref{gasflux} we list the total H$\beta$ and [OIII] fluxes, obtained summing the line fluxes in all the bins that meet the 
A/N criterion. There are generally no major 
differences between the velocity fields in the two lines, while the velocity dispersion is usually higher in H$\beta$; exceptions to this statement
will be highlighted in the individual galaxies description (Section \ref{notes}). The differences between the two lines are 
particularly strong in the case of NGC4102 (see also Figure \ref{bin_bad}
and the related text), which is classified as a LINER. Overall, the sample galaxies display low [OIII]/H$\beta$ ratios over most of the {\tt SAURON} field, suggesting star formation, as 
expected in spiral galaxies.

\subsection{Notes on individual galaxies}\label{notes}
Here we comment individually on the galaxies of our sample; for each of them, we first summarize the known properties from previous work, then we describe
the most interesting features seen in our {\tt SAURON} maps, shown in Section \ref{allthemaps}.

\subsubsection{NGC488}\label{n488}
NGC488 is the most ``elliptical-like'' galaxy in our sample; it is an unbarred Sb galaxy with a very regular tightly wound spiral pattern, with 
a pitch angle of only $5^\circ$ (\citealt{kormendy}, Gerssen, Kuijken \& Merrifield 1997), and a large and smooth central bulge. According to \citet{marcella97}, this is a
classical $r^{1/4}$ bulge. The rotation curve has been measured using H$\alpha$ 
and [NII] 6583\AA\, data by \citet{peterson} 
out to a radius of 20 
kpc, where he found that the velocity reaches $\approx$ 363 km s$^{-1}$, and continues to increase outward. This galaxy has been studied using photometric 
and kinematic data to construct a dynamical model of bulge and disc \citep{fuchs}, concluding that the disc is dynamically
cold.\\
\indent The {\tt SAURON} maps show regular stellar and gaseous rotation, with stars and gas rotating fast around the photometric minor axis and stellar velocity 
 dispersion smoothly increasing 
 toward the centre (consistent with the presence of the large bulge), to reach the highest value in our sample (around 200 km s$^{-1}$, see also Table \ref{properties} and Figure
 \ref{sigma_gradients}). The gas turns out to be quite patchy (H$\beta$) and centrally concentrated.  

\subsubsection{NGC628} \label{n628}
NGC628, also known as M74, is a well-studied grand design spiral galaxy classified as Sc. It is known to be surrounded by an elliptical 
ring of neutral hydrogen 
extending well beyond the optical disc, out to $\approx$ 3 $R_{Ho}$ (isophotal radius at the 26.5 mag arcsec$^{-2}$ surface brightness level), lying in a plane which is 
$\approx$ 15$^\circ$ inclined with respect to the plane of the bright inner disc (see \citealt{rob}, \citealt{briggsold}, \citealt{briggsnew}). The origin of the warped 
velocity field 
of the outer galaxy has been debated, since the apparent isolation of this galaxy makes it difficult for tidal disruption to be responsible for the warp. UV observations 
of NGC628 have shown spiral arms with a more symmetrical appearance than in the optical; \citet{chen} identified in the ultraviolet 
a possible companion 7\farcm6 southwest from the nucleus, which could have helped to clarify the origin of the outer warp, but later it turned out to be a spurious detection. From both NIR spectroscopy of CO absorption and submillimetric imaging 
of CO emission, a circumnuclear ring of star formation has been seen (\citealt{wakker}, \citealt{james}). These rings are believed to exist as a result of
a barred potential; \citet{seigar} claims that ground-based NIR images 
suggest the presence of an oval distortion at $\approx$ 2 kpc from the centre, which could be part of the dissolution of a bar and responsible 
for the circumnuclear ring. On the basis of their analysis of HST $H$-band images, \citet{laine} found a bar on a $\approx$ 100 parsec scale; this would then be a case of nested bars.\\ 
\indent From the stellar population side, \citet{cornett} concluded that the star formation history of NGC628 varies with galactocentric 
distance; Natali, Pedichini \& Righini (1992) suggested that the galaxy could be seen as an inner and an outer disc characterized by different stellar
populations; according to them, the transition between the two regions is located at $\approx$ 8-10 kpc from the centre.\\
 \indent NGC628 can be resolved from the ground into individual supergiants and HII regions (Sohn, Young-Jong \& Davidge 1996). Many previous studies focussed on the HII regions; 
\citet{belley} derived reddenings, H$\beta$ emission equivalent widths and metallicities for 130 HII regions; \citet{ivanov} classified 
a similar number of stellar associations. \citet{davidge} studied a large number (more than 300) of stellar objects in the disc of
NGC628, measuring brightness and colours, finding some recent star formation and getting an estimate of the distance modulus (29.3 mag).\\
 \indent Our observations of NGC628 are disturbed by the presence of a foreground star which falls close to the centre
 ($\approx$ 13$''$ southern); as mentioned in the caption to Table \ref{pointings} our first exposure on this galaxy was
centered on this foreground star. During our last two exposures, the observing conditions were not optimal: the 
seeing went up to $\approx$ 2$''$.\\ 
\indent From our data we can measure slow projected rotation around the minor axis, as expected for an 
almost face-on object. The stellar velocity dispersion decreases in the central zones, indicating a cold central region, maybe an inner disc. Photometric observations at 
UV wavelengths from \citet{cornett} did not show evidence for a bulge: the nucleus of NGC628 has the appearance of disc
material in the UV. The ionised gas rotates in a way 
similar to the stars; the H$\beta$ distribution turns out to be more extended than the 
[OIII] distribution, both of them suggesting an annular structure. 

\subsubsection{NGC772}\label{n772}
NGC772, named also Arp78, forms a pair at 3\farcm3 with the E3 galaxy NGC770. It is an Sb galaxy characterised by a particularly strong 
spiral arm; it is known to have faint HII regions \citep{oye} and extended HI, to a radius $\approx$ 75 kpc \citep{rao}.\\
\indent The unsharp-masked {\tt SAURON} image (no optical HST image is available) displays a well defined dust pattern, with the dust 
following the spiral arms. From the stellar kinematical maps we can see that in the central region there is a clear drop in 
the stellar velocity dispersion 
and $h_{3}$ turns out to anticorrelate with the velocity, as expected for a
rotating disc; the stellar velocity map shows rotation around the photometric minor axis and a mildly S-shaped zero-velocity line. 
The gas, especially in H$\beta$, rotates in a more complex way than the stars, with a very strongly S-shaped 
zero-velocity line, and its distribution follows the spiral arm pattern. The [OIII] velocity field is instead much more regular. The gas velocity dispersion in the H$\beta$ line is flat -and low- 
beyond $\approx$ 10$''$, enhanced in an annular region around the centre and depressed again in the nucleus -although still well higher than in the outer parts. All of this is also seen in the [OIII]/H$\beta$ map, 
where we can recognize regions with very low values corresponding to the spiral arms and to low values in the ionised-gas velocity dispersion. 

\subsubsection{NGC864}\label{n864}
Nuclear radio emission has been found in this barred Sc spiral galaxy \citep{ulverstad};  
the nuclear radio source has diffuse morphology, with linear size of 
$\approx$ 300 pc. \\
\indent The {\tt SAURON} maps show rotation both in the stellar and gaseous components, around the photometric major axis, e.g. the rotation axis is oriented as the bar; but looking at a 
larger scale image (see for example Figure \ref{sample}) we can see that the major axis of the bar turns out to be the
global minor axis when considering the whole optical 
galaxy. The stellar velocity dispersion is flat and low over the entire {\tt SAURON} FoV. The gas flux follows the nucleus + bar structure, as visible especially in the H$\beta$ case. 
The line ratio [OIII]/H$\beta$ is also structured in a similar way, 
assuming low values along the bar and in particular in the central $\approx$ 4$''$ circle, which could indicate a star forming nucleus. Interestingly, this region
corresponds to a local minimum in the H$\beta$ velocity dispersion map, which assumes in the centre lower values 
than the dispersion in [OIII], differently to what happens in most of our galaxies.  

\subsubsection{NGC1042}\label{n1042}
This Scd galaxy forms a pair with the Sc galaxy NGC1035 at a separation of 22$'$ (corresponding to 177 kpc). It has a bright, small
nucleus and otherwise low surface brightness. Neutral hydrogen has been detected at positions corresponding to the optical centre 
and two adjacent regions 
\citep{bottinelli}.\\
\indent In our data, this galaxy has a quite poor signal-to-noise, so that the stellar binning is quite heavy. There are some indications of slow projected stellar rotation, but in general the kinematical maps are difficult to
 interpret. The gas is quite patchy and does not cover, at the chosen A/N level, the whole {\tt SAURON} field. In contrast to
 what is seen in the majority of our galaxies,
 the velocity dispersion in [OIII] is higher than in H$\beta$, especially in the central region.

\subsubsection{NGC2805}\label{n2805}
NGC2805 is an Sd galaxy seen nearly face-on and it is the brightest member of a multiple interacting system containing also NGC2814 (Sb), NGC2820 (Sc)
at 13$'$ and IC2458 (I0). According to \citet{hodge}, 
in this group the HII 
regions appear distorted on the side of the galaxy opposite to the companion. HI has been detected (\citealt{reak}, \citealt{bosma}) and there are claims 
that the outer HI layers are warped (see for example \citealt{bosma}). The galaxy seems to be also optically disturbed, since the spiral arms appear to be 
broken up into straight segments.\\
\indent From our {\tt SAURON} data, we find slow projected stellar velocities and a central drop in velocity dispersion. The gas has a clumpy distribution and rotates consistently with the stars. The line ratio [OIII]/H$\beta$ is low all over the field, possibly indicating ongoing star
formation everywhere. 

\subsubsection{NGC2964}\label{n2964}
NGC2964 is a barred Sbc galaxy and forms a non-interacting pair with the I0 galaxy NGC2968 at 5\farcm8. From HST images \citet{marcella97} detected a resolved central compact component, possibly star forming. 
CO has been detected by \citet{braine}; they suggested that the galaxy might contain an unresolved nuclear ring.\\
\indent Our data reveal a variety of features in this galaxy. 
The stars rotate in a quite regular way, although a twist in the zero-velocity line can be seen very close to the edge of the field, on both sides. 
The stellar velocity dispersion increases smoothly towards the centre. Two-dimensional maps of stellar velocity and velocity dispersion for this galaxy have recently
been published by \citet{batcheldor}. Although their maps deliver a lower spatial resolution than ours and a smaller FoV, there is a global agreement between the two
observations. They also provide central values for the higher Gauss-Hermite moments $h_{3}$ and $h_{4}$; the latter value is larger than ours.\\
\indent The ionised gas has a clumpy distribution, which could suggest a spiral arm structure, and complex kinematics in both lines, 
with a very irregular zero-velocity line and with the velocity dispersion peaking in an off-centre region. The velocity dispersion in [OIII] is
enhanced with respect to H$\beta$. Overall, 
the gas motions are consistent with the stellar rotation. In the region of the gas-$\sigma$ peak, the spectra show also complex line profiles: in some bins we find double-peaked lines in the [OIII] 
spectral region (see discussion in Section \ref{methodgas}). It could be that we are seeing the
regular gas motion together with an ionization cone, caused by a central AGN, as is probably also the case in NGC 4102. Up to now, NGC 2964 has not been
classified as an AGN yet. The double-peak line shape characterizes only a few spectra, so we applied our 
standard method which fits single Gaussians to the emission-lines. The [OIII]/H$\beta$ line ratio is small over most of the field, and is 
enhanced in a ring-like structure (see also \citealt{braine}) surrounding an elongated nuclear region where the ratio becomes low again. This could be a 
star forming region, in accordance with \citet{marcella97}.  

\subsubsection{NGC3346}\label{n3346}
This is a barred Scd galaxy, for which only little information is available from the literature.
 The {\tt SAURON} data show regular stellar rotation around an axis slightly misaligned with respect to the 
 direction perpendicular to the bar, with complicated structures in the very inner
regions; the stellar velocity dispersion is everywhere low and flat, and it seems to be depressed in a 
 central region elongated in the same direction as the bar; the map is quite patchy. The ionised gas is concentrated along the bar and the spiral arms and rotates similarly to 
 the stars, although the gas velocity fields appear quite patchy. The H$\beta$ distribution is more extended than [OIII]. The line ratio [OIII]/H$\beta$  is depressed along the bar, particularly 
 in bubble-shaped spots, as the H$\beta$ velocity dispersion does as well; these could be star forming regions. 

\subsubsection{NGC3423}\label{n3423}
Also for this Scd galaxy there is not much literature to refer to. In the {\tt SAURON} maps we measure slow projected stellar rotation and low velocity dispersion. The gas shows a clumpy distribution and rotates faster than the stars and around the same axis. [OIII]/H$\beta$ is everywhere low, 
possibly indicating wide-spread star formation.

\subsubsection{NGC3949}\label{n3949}
NGC3949 is an Sbc galaxy in the Ursa Major cluster. Optical images show a diffuse extended halo \citep{marcv}. It has been observed with the Westerbork 
Synthesis Radio Telescope \citep{marcthesis}: HI is detected well beyond the optical galaxy.\\
\indent The {\tt SAURON} maps show rotation both in the stellar and gaseous components. The stellar velocity dispersion indicates a cold inner
region. This galaxy is part of the sample observed by \citet{batcheldor}; as in the case of NGC2964, their reported central $h_{4}$ 
value is larger than ours. Integral-field observations of NGC3949 are presented also by \citet{westfall}; their central values for the stellar velocity dispersion are in good
agreement with ours; the agreement at larger radii and in the gas velocity dispersion is less satisfactory. The gas has a complex and
clumpy distribution and its velocity fields present the same characteristics as the stellar one: global rotation and 
irregular zero-velocity line. The global appearance of the previously mentioned HI velocity field resembles our {\tt SAURON} velocity maps. 
Overall, [OIII]/H$\beta$ is low, especially in a circumnuclear region. 
No correspondence between the line ratio and the highly complex gas velocity dispersion maps is evident.

\subsubsection{NGC4030}\label{n4030}
\citet{marcella98b} analysed the HST images of NGC4030, and found tightly wound flocculent spiral structure reaching the nucleus and an irregular,
exponential bulge. 
Our observations of NGC4030 have a very high signal-to-noise, so that the data are of extremely good quality. The kinematical behaviour of this Sbc galaxy is
characterized by very regular rotation of stars and gas 
around the minor axis. The velocity dispersion increases smoothly moving inwards, becoming possibly flat in the nuclear region (see 
Figure \ref{sigma_gradients}), and the stellar $h_{3}$ anticorrelates 
with the velocity, as expected for a rotating disc. Our gas flux maps show a central concentration, although in H$\beta$ some structure is visible that could be related to the spiral arms and/or dust lanes
pattern: as can be seen in the unsharp-masked image and as already noticed by \citet{marcella98b}, the spiral arms and dust lanes structures extend down to the innermost
scales. From long-slit H$\alpha$ optical spectroscopy extended over $\approx$ 80\% of the optical image,
 \citet{mathew} derive a rotation curve with maximum rotation velocity of $\approx$ 236 km s$^{-1}$. Our gas velocity regularly increases going outward, up to a value of 
 $\approx$ 180  km s$^{-1}$ at the edges of the field. Since the outermost radius of our data is smaller than the one of the quoted observations, there is not necessarily a disagreement. 

\subsubsection{NGC4102}\label{ngc4102}
NGC4102 shows a bright central bar from which two tightly wound and dusty spiral arms depart \citep{marcthesis}. In the NED database this object is classified as a LINER; 
it is known to be a powerful far-infrared galaxy \citep{young} and also to have a strong nuclear radio source
\citep{condon}. \citet{devereux} classified it as 
one of the most powerful 
nearby starburst galaxies. From optical spectroscopy of the ionised swept-up gas, \citet{boer} determined the timescale of the central starburst wind 
to be of the order of 10$^{6}$ years: a young starburst. Gon\c{c}alves, V\'eron-Cetty \&
V\'eron (1999) recognized a weak Seyfert2 component, mainly detected by the broadening of the [OIII] lines. NGC4102 
has also been observed
 in CO: \citet{jogee} found a compact CO morphology and a CO velocity field characterized in the inner 200 pc by purely circular motion, and a sharp discontinuity in gas kinematics at larger radii, 
 around 3$''$; this could be due to streaming motions along the bar. HI has been detected in absorption against the bright central radio source \citep{marcthesis}.\\
\indent The signs of all the documentated activity going on in this galaxy are detectable also in our {\tt SAURON} maps: the gas kinematics turn out to be extremely complicated. As mentioned 
in the discussion in Section \ref{gaskinobs}, this is the only case in our sample with a significant difference between the kinematics of the H$\beta$ and [OIII] lines: the 
H$\beta$ velocity field resembles more closely the stellar one, although it is more irregular, while the [OIII] maps trace more likely the outflowing gas. This can be also seen by looking at the 
gas flux maps: in [OIII] the central gaseous concentration appears to be elongated in a northwestern direction, which is not the case in the  H$\beta$ distribution, 
that follows instead the spiral arm pattern. The [OIII] velocity 
dispersion is also elongated in the same direction. Together with NGC2964, this is the 
only galaxy in our sample where we find emission-lines that are not well fitted by single Gaussians (see the discussion in Section \ref{methodgas} and Figure
\ref{bin_bad}). The complex line profiles are found only in the region corresponding to the maximum difference between the H$\beta$ and [OIII] kinematics (which is also, as
already noticed, the region of the [OIII] $\sigma$ peak and flux elongation); there the [OIII] lines turn out to be broadened, in accordance with \citet{goncalves}.
According to our measurements, the [OIII]/H$\beta$ line ratio is then enhanced in the outflow region, as expected for an active object. As for the stellar kinematics, 
we do not see signs of activity there: the stars behave in a much more quiet way, with regular rotation and velocity dispersion increasing from the outer parts toward the centre. 

\subsubsection{NGC4254}\label{n4254}
NGC4254, known also as M99, is one of the brightest spiral galaxies in the Virgo cluster. Its optical appearance is characterised by the one-arm structure: 
the arms to the northwest are much less defined than the southern arm. This kind of spiral structure could be 
related to an external driving mechanism, but for NGC4254 there are no close companions and no evident signs of past interactions \citep{rauscher}. Phookun, 
Vogel \& Mundy (1993) carried out 
HI and H$\alpha$ observations of this Sc galaxy, detecting non-disc HI emission, coming mostly from a region to the north of the galaxy
and contiguous with it in the plane of the sky; according to the authors, the non-disc HI corresponds to 3\% of the total HI mass; 
they do not find H$\alpha$ emission corresponding to the non-disc gas, which indicates little or no star formation there. They interpreted these observational results as 
infall of a disintegrating gas cloud; this could have caused the asymmetry in the spiral structure. H$\alpha$ intensity and velocity fields 
are presented also in \citet{chemin2005}, confirming the asymmetry in the prominent spiral structure and detecting streaming motions along the
arms. There are also claims for the presence of a weak bar with a PA of
$\approx$ 60$^\circ$, although NGC4254 is classified as unbarred (\citealt{sakamoto}, from analysis of the molecular gas distribution).\\
 \indent We observed this galaxy under bad seeing conditions ($\approx$ 3$''$). Our kinematic maps reveal regular stellar rotation, although 
 the stellar zero-velocity line is twisting toward the edges of the field. 
 The gas flux follows the spiral pattern, which is particularly evident in the H$\beta$ case and can be seen also in 
the [OIII]/H$\beta$ line ratio, which assumes particularly low values in bubble-shaped regions along the arms. The gas rotation globally
resembles the stellar motions, but presents more complex structures, with S-shaped zero-velocity lines, which could be related to the mentioned possible bar; Fourier analysis of the velocity field will help clarifying this.
 Velocity dispersion in [OIII] is higher than in H$\beta$, as particulary visible in the central region. 

\subsubsection{NGC4487}\label{n4487}
This Scd galaxy forms a pair with the Scd galaxy NGC4504, at a separation of 35$'$ (corresponding to $\approx$ 
165 kpc). Two principal arms of the grand-design type 
are recognizable in the images; one of the two is less well defined than the other and splits into broad segments which cover one side of the disc. The spiral pattern shows 
up also in our gas flux maps. The stars and gas rotate around an axis misaligned with respect to the photometric minor axis. 
The stellar and gaseous velocity dispersions are everywhere very low. In the [OIII]/H$\beta$ map regions of possible star formation can be recognized, in the centre and along the spiral 
arm, in correspondence to a drop in the H$\beta$ velocity dispersion. Contrary to what happens in the majority of our galaxies, 
in the very centre the velocity dispersion in 
H$\beta$ is lower is than in [OIII].

\subsubsection{NGC4775}\label{n4775}
Very little is known about this low-inclination and very late-type (Sd) galaxy. Our maps, although based on 4 $\times$ 1800s exposures, 
are quite patchy and not easy to interpret; they suggest very slow stellar and gaseous projected rotation, as expected for an almost face-on galaxy, 
low stellar and gaseous velocity dispersion
and a clumpy gas distribution in both H$\beta$ and [OIII] lines.  

\subsubsection{NGC5585}\label{n5585}
NGC5585 is a barred Sd galaxy; together with NGC5204 (Sm), NGC5474 (Scd), NGC5477 (Sm) , HoIV (Im) and M101 (Scd) it forms the M101 group. It is highly resolved into individual stars and HII regions. 
In this galaxy Cot\'e, Carignan \& Sancisi (1991) detected HI extended out to more than twice the optical radius; the HI velocity field
turned out to be strongly warped; using these data and the $B$ band 
luminosity profiles, they constructed a mass model finding that the contribution of the dark matter component dominates the rotation curve at almost all radii, also when using a 
maximum-disc method. Later, \cite{blaise} pointed out that using two-dimensional H$\alpha$ Fabry-P\'erot spectroscopy one can better constrain the orientation
parameters and the rotation curve in its rising part, reducing by 30\% the dark to luminous matter ratio.\\
\indent In our {\tt SAURON} maps there is not much information about the stellar velocity (patchy map). There are some hints of slow projected 
rotation. The stellar $\sigma$ decreases
moving outwards from the central region, although the map appears patchy and has everywhere low values. The gas distribution follows the bar morphology; along the bar some spots can be seen where the flux 
in [OIII] is larger than in H$\beta$ . The [OIII]/H$\beta$ line ratio is enhanced in a region 
elongated in a roughly perpendicular direction to the one defined by the bar.    

\subsubsection{NGC5668}\label{n5668}
NGC5668 is another Sd galaxy, which has a high rate of star formation, as indicated by its large far-infrared and 
H$\alpha$ luminosities (\citealt{schulman}, and references therein); the mentioned 
authors observed it in HI and detected high-velocity clouds of neutral hydrogen. High velocity clouds 
were found also in the ionised gas and interpreted as regions with 
vertical motions related to ongoing star forming processes in the disc, as pointed out by \citet{jimenez} on the
basis of Fabry-P\'erot 
H$\alpha$ observations.\\
\indent Our {\tt SAURON} data indicate very slow stellar projected rotation velocities and a cold inner region.  The gas distribution is clumpy, defining a ring-like structure also characterized by slightly lower 
[OIII]/H$\beta$ values compared to the surroundings.

\subsubsection{NGC5678}\label{n5678}
Not much is known from previous work on this barred and very dusty Sb galaxy. According to our {\tt SAURON} data,   
the stars rotate in a rather regular way around the minor axis, although the zero-velocity line is strongly bent. 
The stellar velocity dispersion increases moving inwards, and becomes flatter in the nuclear region (see Figure \ref{sigma_gradients}). The gas 
has instead a complex distribution, more structured in H$\beta$ than in [OIII], with kinematics consistent with the stars, but a zero-velocity line 
more disturbed and wiggling, similar to what we see in NGC2964. Over a large part of the field, the velocity dispersion in the [OIII] line is higher
than in H$\beta$.

\section{CONCLUSIONS}\label{conclusionsec}
Two-dimensional kinematics and stellar population analysis of spiral galaxies toward the end of the Hubble sequence (Hubble types later
than Sb) is still a relatively 
unexplored field: late-type spirals are very complex objects, often faint and full of substructures, as recently proved by
analysis of HST images. They have been the target of a few 
photometric and long-slit optical spectroscopic observations, but measurements of their two-dimensional kinematics were still missing. 
We have started a project on a sample of 18 such objects using integral-field spectroscopic observations obtained with {\tt
SAURON}. This allowed us 
to measure the stellar kinematics, the flux and kinematics of the H$\beta$ 4861\AA\, and [OIII]$\lambda\lambda$4959,5007\AA\, emission-lines and the strength of 
the H$\beta$, Fe and Mg{\textit{b}} absorption features over a two-dimensional area covering the central region of our galaxies.\\
\indent In this paper we discussed the first results from this study, presenting the two-dimensional kinematics for stars and ionised gas. The majority of our galaxies is shown to be 
kinematically cold and to possess 
a considerable amount of ionised gas, covering in most cases a large part of the {\tt SAURON} FoV and frequently following bar or spiral arm patterns in the 
spatial distribution. A quite common feature of our measured stellar kinematical maps is a central depression in the velocity
dispersion, which assumes very often low values; we measured the velocity dispersion profiles and correlated their slopes with the morphological type: later-type galaxies tend to have velocity dispersion profiles 
which increase outwards. This implies small bulge/disc ratios and the presence of inner, occasionally star-forming, 
disc-like structures. We also qualitatively compared the characteristics of our maps with the galaxy's properties known from literature HST isophotal analysis: the main 
conclusion common to spectroscopy and photometry is that the kinematic detection of a cold inner region turns out to be often related to the lack of a classical stellar 
bulge and the presence of small-scale structures (nuclear star clusters, inner rings, inner bars). The 
gaseous component turns out to be almost ubiquitous and kinematically highly complex, displaying in many cases irregular velocity fields, with the kinematic axis twisting or 
bending or wiggling, or even without a clear sense of rotation, possibly because of the dust which strongly affects these objects. They also host intense star
formation, often spread over the whole region we have observed, as suggested by the low values in the [OIII]/H$\beta$ line ratio maps.\\
\indent In follow-up papers we will model the observed kinematic fields in detail, present the line-strength maps for 
these same galaxies, consider the bulge-disc decomposition, and compare our
results with those for the 24 Sa bulges in the {\tt SAURON} survey. 

\section*{ACKNOWLEDGEMENTS}
We kindly acknowledge Anna Pasquali and Ignacio Ferreras for their help during the early
stages of this project. We thank Marc Balcells, Marc Sarzi and Kristen Shapiro for a careful reading of and commenting on the draft. We also thank 
the referee Torsten B\"oker for useful comments and suggestions which improved the paper. 
 The {\tt SAURON} - related projects are made possible through grants 614.13.003, 781.74.203, 614.000.301 and 614.031.015 from NWO and financial contributions from the Institut 
 National des Sciences de l'Univers, the Universit\'e Claude Bernard Lyon~I, the Universities of Durham, Leiden and the Netherlands 
 Research School for Astronomy NOVA. KG acknowledges support for the Ubbo Emmius PhD program of the University of Groningen. JFB acknowledges support from the Euro3D 
 Research Training Network, funded by the EC under contract HPRN-CT-2002-00305. MC acknowledges support from a VENI
 grant 639.041.203 awarded by the Netherlands Organization for Scientific Research (NWO). This project made use of the HyperLeda and NED databases. Part of this work is based 
 on data obtained from the STSci Science Archive Facility. The Digitized Sky Surveys were produced at the Space
 Telescope Science Institute under U.S. Government grant NAGW-2166. The images of these surveys are based on
 photographic data obtained using the Oschin Schmidt Telescope on Palomar Mountain and the UK Schmidt Telescope.

%%%%%%%%%%%%%%%%%%%%%%%%%%%%%%%%%%%%

\makeatletter
\def\thebiblio#1{%
 \list{}{\usecounter{dummy}%
         \labelwidth\z@
         \leftmargin 1.5em
         \itemsep \z@
         \itemindent-\leftmargin}
 \reset@font\small
 \parindent\z@
 \parskip\z@ plus .1pt\relax
 \def\newblock{\hskip .11em plus .33em minus .07em}
 \sloppy\clubpenalty4000\widowpenalty4000
 \sfcode`\.=1000\relax
}
\let\endthebiblio=\endlist
\makeatother

%%%%%%%%%%%%%%%%%%%%%%%%%%%%%%%%%%%%

\label{lastpage}

\end{document}